\pdfoutput=1
\documentclass[a4paper,fleqn]{cas-dc}
\usepackage[numbers]{natbib}
\usepackage{paralist}
\usepackage{graphicx}
\usepackage{textcomp}
\usepackage{xcolor}
\usepackage{url}
\usepackage{multirow}
\usepackage{hyperref}
\usepackage[listings]{tcolorbox} 
\usepackage{subcaption}
\usepackage{soul}
\usepackage[normalem]{ulem}
\usepackage{balance}
\usepackage{framed}
\usepackage{amsmath,amssymb,amsfonts}
\usepackage{algorithmic}
\usepackage{array}
\usepackage{threeparttable}

\newcolumntype{R}[1]{>{\raggedleft\arraybackslash }m{#1}}
\newcolumntype{L}[1]{>{\raggedright\arraybackslash }m{#1}}
\newcolumntype{C}[1]{>{\centering\arraybackslash }m{#1}}

\newcommand{\yun}[1]{\textcolor{black}{{#1}}}
\newcommand{\add}[1]{\textcolor{black}{{#1}}}
\newcommand{\mystar}{{\fontfamily{lmr}\selectfont$\star$}}

\makeatletter
\newcommand{\thickhline}{%
	\noalign {\ifnum 0=`}\fi \hrule height 2pt
	\futurelet \reserved@a \@xhline
}
\newcolumntype{"}{@{\hskip\tabcolsep\vrule width 1pt\hskip\tabcolsep}}
\makeatother

\def\tsc#1{\csdef{#1}{\textsc{\lowercase{#1}}\xspace}}
\tsc{WGM}
\tsc{QE}
\tsc{EP}
\tsc{PMS}
\tsc{BEC}
\tsc{DE}

\begin{document}
\sloppy
\let\WriteBookmarks\relax
\def\floatpagepagefraction{1}
\def\textpagefraction{.001}
\shorttitle{Do Users Care about Ad's Performance Costs?}
\shortauthors{Cuiyun Gao et~al.}

\title [mode = title]{Do Users Care about Ad's Performance Costs? Exploring the Effects of the Performance Costs of In-App Ads on User Experience}                      



\author[1,2]{Cuiyun Gao}
\ead{cygao@cse.cuhk.edu.hk}

\address[1]{School of Computer Science and Technology, Harbin Institute of Technology, Shenzhen, China}
\address[2]{Department of Computer Science and Engineering, The Chinese University of Hong Kong, Hong Kong, China}

\author[2]{Jichuan Zeng}
\corref{cor1}
\ead{jczeng@cse.cuhk.edu.hk}

\author[3]{Federica Sarro}
\ead{f.sarro@ucl.ac.uk}
\address[3]{Department of Computer Science, University College London, United Kingdom}

\author[4]{David Lo}
\ead{davidlo@smu.edu.sg}
\address[4]{School of Information Systems, Singapore Management University, Singapore}

\author[2]{Irwin King}
\ead{king@cse.cuhk.edu.hk}

\author[2]{Michael R. Lyu}
\ead{lyu@cse.cuhk.edu.hk}


\nonumnote{* J. Zeng is the corresponding author}


\begin{abstract}
\textit{Context:} In-app advertising is the primary source of revenue for many mobile apps. The cost of advertising (ad cost) is non-negligible for app developers to ensure a good user experience and continuous profits. Previous studies mainly focus on addressing the hidden performance costs generated by ads, including consumption of memory, CPU, data traffic, and battery. However, there is no research on analyzing users' perceptions of ads' performance costs to our knowledge.

\noindent\textit{Objective:} To fill this gap and better understand the effects of performance costs of in-app ads on user experience, we conduct a study on analyzing user concerns about ads' performance costs.

\noindent\textit{Method:} First, we propose RankMiner, an approach to quantify user concerns about \yun{specific} app issues, \yun{including performance costs}. Then, based on the usage traces of 20 subject apps, we measure the performance costs of ads. Finally, we conduct correlation analysis on the performance costs and quantified user concerns to explore whether users complain more for higher performance costs.

\noindent\textit{Results:} Our findings include the following: (1) RankMiner can quantify users' concerns better than baselines by an improvement of 214\% and 2.5\% in terms of Pearson correlation coefficient (a metric for computing correlations between two variables) and NDCG score (a metric for computing accuracy in prioritizing issues), respectively. (2) The performance costs of the with-ads versions are statistically significantly larger than those of no-ads versions with negligible effect size; (3) Users are more concerned about the battery costs of ads, and tend to be insensitive to ads' data traffic costs.

\noindent\textit{Conclusion:} Our study is complementary to previous work on in-app ads, and can encourage developers to pay more attention to alleviating the most user-concerned performance costs, such as battery cost.
\end{abstract}




\maketitle

\section{Introduction}\label{sec:intro}
In-app advertising is a type of advertisement (ad) within mobile applications (apps). Many organizations have successfully monetized their apps with ads and reaped huge profits. For example, the mobile ad revenue accounted for 76\% of Facebook's total sales in the first quarter of 2016~\citep{facebook}, and increased 49\% year on year to about \$10.14 billion in 2017~\citep{fbadincrease}. Triggered by such tangible profits, mobile advertising has experienced tremendous growth recently~\citep{adreport}. Many free apps, which occupy more than 68\% of the over two million apps in Google Play~\citep{market}, adopt in-app advertising for monetization. However, the adoption of ads has strong implications for both users and app developers. For example, almost 50\% of users uninstall apps just because of ``intrusive'' mobile ads~\citep{adsurvey}, which may result in a reduction in user volume of the apps. Smaller audiences generate fewer impressions (\textit{i.e.}, ad displaying) and clicks, thereby making ad profits harder for developers to earn. Thus, understanding the effects of in-app ads on user experience is helpful for app developers.

User reviews serve as an essential channel between users and developers, delivering users' instant feelings (including the unfavorable app functionalities or annoying bugs) based on their experience. User review mining has been proven useful and significant in various aspects of app development, such as supporting app design~\cite{DBLP:conf/kbse/GuK15}, categorizing app issues for facilitating app maintenance~\cite{Palomba2017ICSE,DBLP:conf/re/MaalejN15}, and assisting app testing~\cite{Grano2018}, etc. In this work, we resort to user reviews to identify users' perception about in-app ads.

Previous research has been devoted to investigating the hidden costs of ads, \textit{e.g.}, energy~\citep{mohan2013prefetching}, traffic~\citep{nath2015madscope}, system design~\citep{grace2012unsafe}, maintenance efforts~\citep{gui2015truth}, and privacy~\citep{sooel2016what}. For example, Gui \textit{et al.}~\citep{gui2015truth} found that the with-ads apps can lead to 30\% more energy consumption than the corresponding no-ads versions. \yun{Relieving all the types of hidden costs is quite labor-intensive for app developers. Understanding users' concerns about these costs can help developers focus on the user-concerned cost types and reduce labor cost.}
Although there are studies using surveys to understand users' perceptions of mobile advertising, \textit{e.g.}, interactivity~\citep{yu2013you}, perceived usefulness~\citep{soroa2010factors}, and credibility~\citep{chowdhury2006consumer}, there is still a lack of study on analyzing users' concerns about the practical performance costs of in-app ads. There are several challenges to perform this kind of analysis. First, collecting a large amount of user feedback that reflects ads' performance costs is intractable. According to Gui \textit{et al.}'s manual analysis of 400 sample ad-reviews~\cite{gui2017arx}, only four (1\%) of the reviews are related to mobile speed and one (0.25\%) relates to battery. Moreover, only around 1\% of collected reviews clearly deal with in-app ads. Second, users' concerns about ad costs are difficult to be quantified, where user behaviors (such as rating apps) should be well involved. Lastly, measuring the performance costs solely incurred by ads is difficult practically due to diverse usage patterns (\textit{e.g.}, different ad viewing duration) from users.

In this paper, we try to overcome these challenges, and propose an approach, named RankMiner, to \yun{quantitatively measure user concern levels} about specific app issues. \yun{Note that RankMiner can measure users' concerns about any specified app issues besides the performance-related ones studied in this paper.} To verify the effectiveness of RankMiner, we choose CrossMiner~\cite{man2016experience}, an issue ranking framework, as one baseline. Experiments show that our approach can outperform baselines by up to 2.5\% in NDCG score~\cite{croft2010search} (a metric for computing accuracy in prioritizing issues) and 214\% in Pearson correlation coefficient~\cite{pearson} (a metric for computing correlations between two variables).

To measure the performance costs incurred by ads practically, we collect usage traces of 17 volunteer users for 20 Android apps containing ads. We focus on measuring four performance cost types: memory consumption, CPU utilization, network usage, and battery drainage, since these costs are commonly discussed in previous studies~\cite{vallina2012breaking, gui2015truth}. The recorded usage traces were then replayed multiple times for simulating real usage scenarios and accurate cost measurement, resulting in the collection of more than 2,000 measurements for those apps. We measure the performance costs of ads based on these measurements.


We focus on answering the following questions:

\begin{inparaenum}[\bfseries RQ1.]
    \item Do the performance costs of in-app ads significantly increase the no-ads versions? We re-analyze some of the questions (e.g., what is the energy cost of ads?) investigated by Gui \textit{et al.}~\cite{gui2015truth} by using practical usage traces for each subject app, whereas Gui \textit{et al.}~\cite{gui2015truth} use one experimental usage trace per app. This allows us to answer this question in a more realistic scenario.\par 
    \item How can the performance costs of ads affect user opinions? Based on the measured performance costs of ads, we empirically analyze the correlations between the costs and the user concerns quantified by RankMiner. We aim at exploring whether users pay more attention to performance costs. \par
\end{inparaenum}

The key \textit{\textbf{contributions}} of this paper are as follows:

\begin{inparaenum}[\itshape \upshape(a\upshape)]
	\item We revisit some questions posed in previous work~\cite{gui2015truth} by using practical usage traces. We find that performance costs of with-ads versions are significantly larger than those of no-ads versions with negligible effect sizes.\par
	\item We carry out the first empirical study to explore correlations between the performance costs of ads and their impact on user opinions, from which we deduce which cost types users care more about. We find that users are more concerned about the battery costs of ads, and tend to be insensitive to ads' data traffic costs.\par
	\item We make the source code\footnote{https://remine-lab.github.io/adbetter.html} of the tools used to measure performance cost and to perform user review analysis publicly available to allow for replication and extension of our work. \par 
\end{inparaenum}

\textbf{Paper structure.} Section~\ref{sec:background} describes the background and motivation of our work. Section~\ref{sec:approach} presents the methodology we propose for quantifying user concerns about specific app issues. Section~\ref{sec:experiment} describes the results of our study. 
Section~\ref{sec:limitation} discusses its limitation.
Related work and final remarks are discussed in Section~\ref{sec:literature} and  Section~\ref{sec:conclusion}, respectively.
\section{Background}\label{sec:background}
In this section, we explain the concept of user reviews, the mobile advertising profit model, and the word embedding technique we utilize in app issue ranking.

\subsection{User Reviews}
\begin{figure}
    \centering
    \includegraphics[width=0.4 \textwidth]{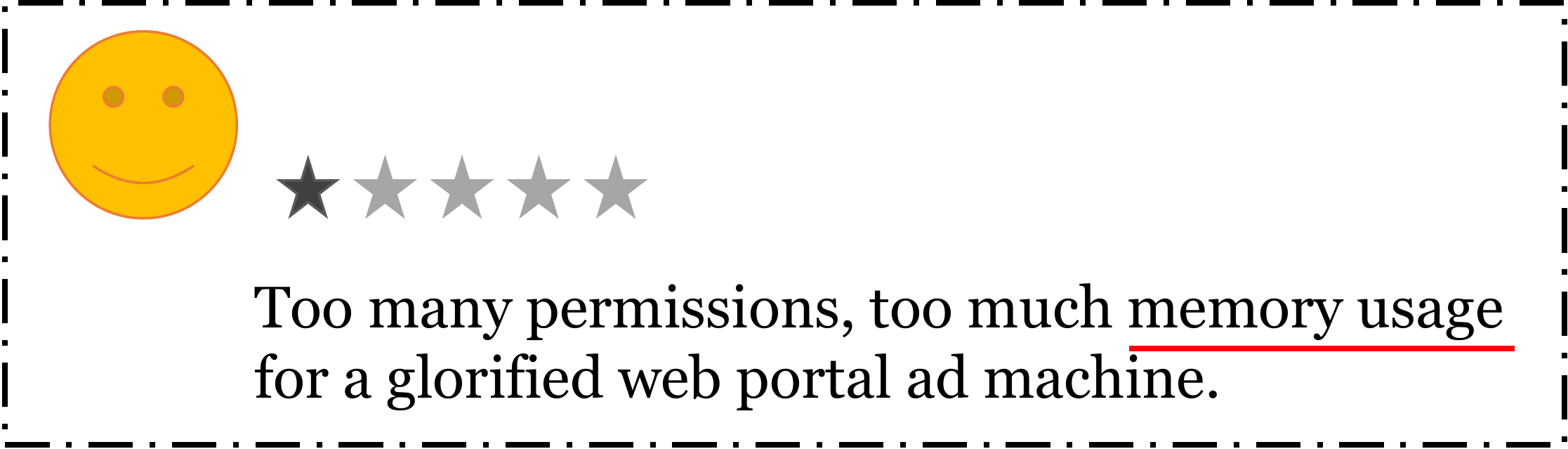}
    \caption{Example of user reviews. \yun{The red underline highlights one 2-gram term (``memory usage'')}.}
    \label{fig:reviewexample}
\end{figure}

User reviews on app distribution platforms, \textit{e.g.}, Google Play, are posted by users to express their experience with apps. They generally serve as the primary channel for customers to leave feedback. As observed~\citep{reviewimportant}, two thirds of users leave reviews after negative experiences. The reviews, which usually report bugs, feature requests, and functionality improvement, provide valuable information to developers.
Figure~\ref{fig:reviewexample} depicts a review of an app publicly available from Google Play (The user's name and the date of the comment have been removed to preserve privacy). The user review complains about memory issues, \yun{indicated by the term ``\textit{memory usage}''}.
Such information can be exploited by developers to discover user experience and improve app design accordingly. More importantly, reviews reflect real and immediate user response after interacting with apps, which cannot be easily collected by surveys. Thus, we leverage app reviews to capture user perceptions of in-app ads in this paper.

\subsection{The In-App Advertising Ecosystem}

\begin{figure}
	\centering
	\includegraphics[width=0.4 \textwidth]{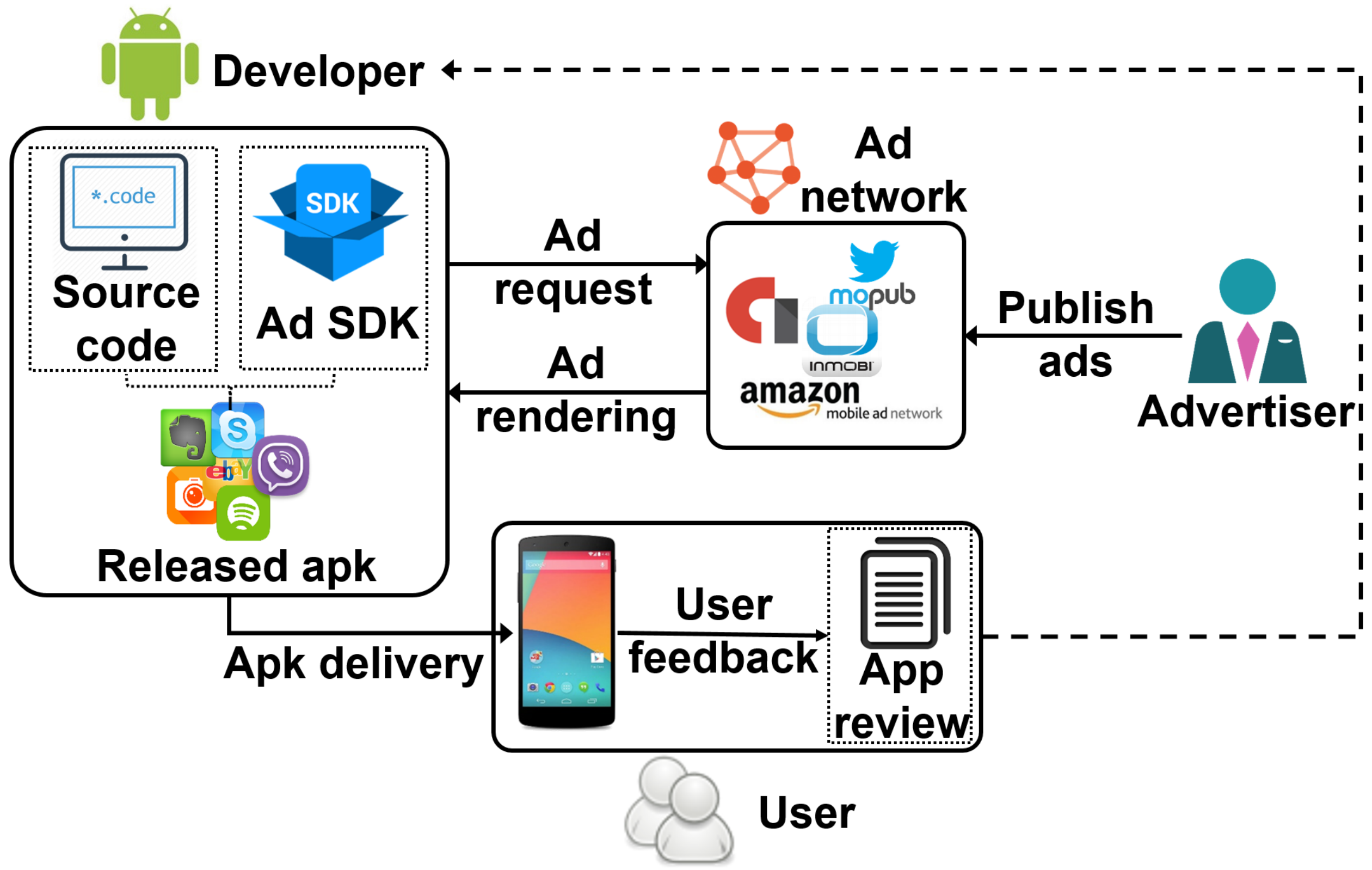}
	\caption{The in-app advertising ecosystem.}
	\label{fig:adflow}
\end{figure}
The ecosystem for in-app advertising consists of four major ingredients, \textit{i.e.}, app developers, advertisers, ad networks, and end users, as shown in Figure~\ref{fig:adflow}. To render advertising contents into an app, developers typically utilize third-party mobile ad SDKs which are provided by ad networks, such as AdMob~\citep{admob} and MoPub~\citep{mopub}. The ad networks grant developers with options for ad display, \textit{e.g.}, defining ad sizes. When loading a page embedded with ads, the app sends a request to the ad network for retrieving an ad. Finally, the fetched ad content is rendered on the user's screen. Developers can then get payment from advertisers according to the counts of ads displayed (\textit{i.e.}, impressions) and clicked.

Users play an essential role in the ad-profiting process, since the number of ads displayed to users determines the ad revenue: User retention and a large user base are critical for app developers. However, embedding ads inappropriately can ruin user experience. According to a survey~\citep{adannoying}, two in three app users consider mobile ads annoying and tend to uninstall those apps or score them lower to convey their bad experience. Such negative feedback likely influences other potential users, which further leads to customer churn. Hence, exploring the effects of in-app ads on user experience is important for app monetization.

\subsection{Word Embedding Techniques}\label{subsec:embedding}
Word embedding also known as word distributed representation~\cite{DBLP:conf/nips/MikolovSCCD13,DBLP:conf/acl/TurianRB10} is a technique for learning vector representations of words by training on a text corpus. Word embedding represents words as fixed-length real-valued vectors so that semantically or syntactically similar words are close to each other in the continuous vector space~\cite{DBLP:conf/nips/MikolovSCCD13}. Word embeddings can be learnt using neural models such as Continuous Bag-of-Words (CBOW) or Skip-Gram~\cite{DBLP:conf/nips/MikolovSCCD13}, where the context words within a sliding window are involved during the learning process. Compared with traditional ``Bag-of-words'' approaches, e.g., counting word frequencies, word embeddings are low-dimensional (often tens or hundreds of dimensions), and thereby do not suffer from sparsity and inefficiency problem.

Likewise, a phrase (i.e., a term with more than one word) can also be embedded as a real-valued vector~\cite{DBLP:journals/tacl/YuD15}. A basic way of phrase embedding is to view it as a bag-of-words and add up all its word vectors.

\section{RankMiner: App Issue Ranking}\label{sec:approach}
\begin{figure}
    \centering
    \includegraphics[width=0.45\textwidth]{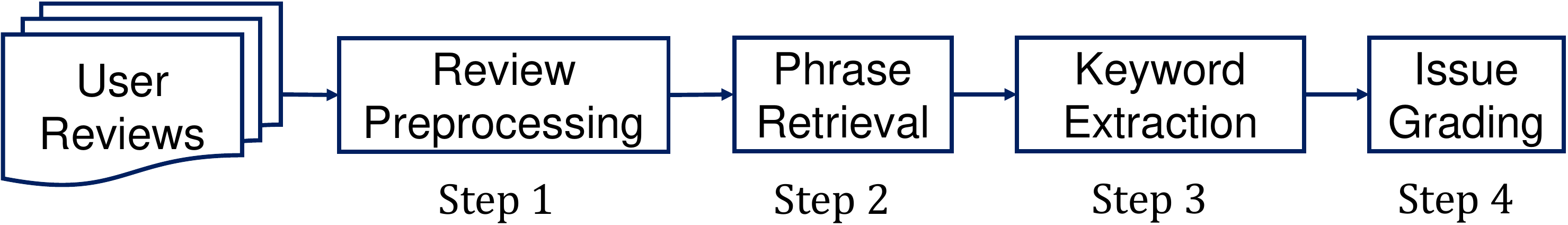}
    \caption{Overview of RankMiner.}
    \label{fig:rankminer}
\end{figure}

Figure~\ref{fig:rankminer} shows the workflow of the proposed approach for quantifying user concerns about specific app issues, which mainly involves four steps: Review preprocessing, phrase retrieval, keyword extraction, and issue grading. We identify phrase candidates from preprocessed reviews, and extract keywords (including phrases and single words) related to specific app issue. Based on the related keyword list, we compute the user concern score about the issue. We elaborate on each step in the following.

\subsection{Preprocessing}\label{subsubsec:preprocessing}
App reviews are usually short in length and contain many casual words. To facilitate subsequent analysis, we eliminate the noisy characters in this step. We first convert all words into lowercase, and remove all non-English characters and non-alpha-numeric symbols. We retain the punctuations to ensure semantic integrity. Then, we reduce the words to their root forms by lemmatization~\citep{nltk}, \textit{e.g.}, ``was'' to ``be''. Finally, we keep reviews with the number of words larger than three. We do not remove stop words~\citep{stopwords} here for phrase retrieval in the next step. Since app reviews contain growing compound words (\textit{e.g.}, redownload), new words (\textit{e.g.}, galaxys8), and misspelled words (\textit{e.g.}, updte [update]), we do not involve the preprocessing methods in~\citep{vu2015mining} where the custom dictionary may introduce information loss (\textit{e.g.}, over correction) in our situation. 

\subsection{Phrase retrieval}\label{subsubsec:phrase}
Phrase retrieval aims to identify the key terms (particularly those with multiple words) that are commonly used by users to voice their experience. The phrases are extracted here since one single word may be ambiguous in its semantic meanings without the context information. For example, in Figure~\ref{fig:reviewexample}, using either ``change'' or ``storage'' alone cannot reflect the comment completely, whereas the three consecutive words ``permit change storage'' can describe the user's viewpoint more accurately.

However, given that user reviews are casually written, extracting the meaningful phrases poses a challenge. In this paper, we adopt the typical Point-wise Mutual Information (PMI) method~\citep{lin2009phrase}. The PMI method measures the co-occurrence probabilities of two words, and thereby eliminating terms which are rarely used. The phrases we retrieve contain 2-gram terms (\textit{i.e.}, two consecutive words) and 3-gram terms (\textit{i.e.}, three consecutive words). Since phrases with more than three words rarely exist in the review collection, they are not extracted here. Equation~(\ref{equ:pmi}) defines the PMI between two words $w_1$ and $w_2$:

\begin{equation}\label{equ:pmi}
\text{PMI}(w_1, w_2) = \log \frac{Pr(w_1\; w_2)}{Pr(w_1)Pr(w_2)},
\end{equation}

\noindent where $Pr(w_1\; w_2)$ and $Pr(w_i)$ denote the occurrence probabilities of the phrase ($w_1\; w_2$) and the single word $w_i$, respectively. The terms with higher PMIs indicate that they appear together more frequently and tend to be semantically meaningful. The PMI thresholds are experimentally set. Based on the PMI results, we also ensure that at least one noun is included in each phrase via the Part-Of-Speech tagging method~\citep{postagging}.

\subsection{Keyword Extraction}\label{subsec:vis}
We propose to use an effective word representation approach, i.e., word emebdding~\citep{mikolov2013efficient} (introduced in Section~\ref{subsec:embedding}), to discover semantically similar words and phrases for each app issue, where the app issue is usually described in keyword, e.g., ``\textit{privacy}'' and ``\textit{crash}''.

We retrieve $k$ terms (including single words and phrases) most close to specific app issues based on the cosine distance of their vector representations, where $k$ is usually defined in the range of tens to hundreds. Due to the small number of the retrieved terms and also to ensure the keyword retrieval accuracy, we then manually trim noise words and phrases. The remaining terms are \textbf{issue-related terms}. 

\begin{table*}
	\small
	\center
	\caption{Example of SentiStrength scores and defined sentiment scores for example review sentences.}
	\label{tab:sentiment}
	\scalebox{0.8}{\begin{tabular}{L{6cm}||C{2cm}||C{2.5cm}}
		\hline
		Review Sentence & SentiStrength Score & Defined Sentiment Score \\
		
	    \hline
	    Great but why make a browser if you don't have the resource to keep it up to date? & [3,-1] & 3 \\
	    \hline
	    Would be 5 stars if i could pay and remove all the ads. & [1,-1] & -1 \\
	    \hline
	    I like what it does but the additional stuff is annoying, eg loud video advert is disturbing. & [2,-3] & -3 \\
	    \hline
	\end{tabular}
}
\end{table*}

\subsection{Issue Grading}\label{issue_detection}
We regard an review related to an app issue if the review containing any terms belonging to the issue-related terms. Similar to previous work~\citep{chen2014ar,wei2017oasis}, we assume that issues complained in more reviews and yielding poorer ratings indicate higher concern levels among users, and need to be ranked higher. The time information (used by Chen et. al~\citep{chen2014ar}) of the issues is not considered here, since we do not care about whether one issue is fresher than the others. In the following, we introduce the sentiment score and frequency score we adopt for grading app issues $\{I_1, I_2, ..., I_i, ..., I_w\}$ where $w$ is the number of app issue to be ranked.

\textbf{Sentiment Score:} The ratings provided by users may not be consistent with the sentiment expressed by their reviews. For example, one user describes ``\textit{Bad app}'' in his/her review but gives a five-star rating. To mitigate this problem, we try to predict the actual sentiment of each user review. Inspired by Guzman and Maalej's work~\citep{guzman2014users}, we use SentiStrengh~\citep{DBLP:journals/jasis/ThelwallBPCK10}, a lexical sentiment extraction tool specialized in dealing with short, low-quality texts, for the sentiment analysis. 


We first chunk the reviews into sentences by utilizing NLTK's punkt tokenizer~\citep{tokenizer}. Then we adopt SentiStrength to assign a positive and negative value to each review sentence, with positive scores in the range of [+1, +5], where +5 denotes an extremely positive sentiment and +1 denotes the absence of sentiment. Similarly, negative sentiments with the range [-5, -1], where -5 denotes an extremely negative sentiment and -1 indicates the absence of any negative sentiment. Table~\ref{tab:sentiment} displays examples of SentiStrength scores for review sentences. If the negative score multiplied by 1.5 is larger than the positive score, the sentence is assigned with the negative score. Otherwise, the final sentiment score is defined as the positive score. As explained in~\citep{guzman2014users}, multiplying the negative scores by 1.5 is considered due to the fact that users tend to write positive reviews~\citep{DBLP:conf/bcshci/IacobVH13}. We finally compute the sentiment score $R_i$ of the issue $I_i$ as the average sentiment scores of all its review sentences.

\textbf{Frequency Score:} The number of the reviews for issue type $I_i$ can be easily calculated, denoted as $N_{i}$.

\textbf{Final Concern Score:} The final concern score $U_{i}$ is defined in Equation~(\ref{eq:userconcern}), by combining the sentiment score $R_i$ and frequency score $N_i$.

\begin{equation}\label{eq:userconcern}
U_{i} = -\log f(R_{i}) \times P_{i},
\end{equation}

\noindent where $P_{i} = N_{i} / N$, representing the percentage of the $I_i$-related reviews in the whole ad-related reviews. The function $f(R_{i})$ is to confine the rating $R_i$ to be in the range $(0,1)$, which is empirically defined as the soft division function, \textit{i.e.}, $(R_{i}-0.9)/5$, or the sigmod function, \textit{i.e.}, $1/(1+e^{-R_{i}})$. Equation~(\ref{eq:userconcern}) shows that issues with lower user ratings and larger review percentages will be given higher user concern values, which is consistent with our initial assumption.

\section{Experimental Study}\label{sec:experiment}
In this section, we first verify the effectiveness of RankMiner as a \add{proof of concept}, and then answer the following research questions as outlined in the Introduction:

\begin{inparaenum}[\bfseries RQ1.]
    \item \yun{Do the performance costs of in-app ads significantly increase the no-ads versions?}\par
    \item \yun{How can the performance costs of ads affect user opinions?} \par
\end{inparaenum}

\subsection{\add{Proof of Concept}: What is the accuracy of the proposed RankMiner approach?}
\subsubsection{Motivation} By answering this question, we aim at verifying the effectiveness of RankMiner in quantifying user concerns about specific app issues. In this way, we can effectively measure user opinions about the performance costs of in-app ads based on RankMiner.

\subsubsection{Methodology} We conduct evaluation of our proposed strategy based on the reviews of Spotify Music provided by CrossMiner~\cite{man2016experience}. We employ this dataset due to its large number of app reviews and available ground truth (\textit{i.e.}, the official user forum~\citep{spotify}). The reviews are collected from three platforms: Android (178,477 reviews), iOS (249,212 reviews), and Windows Phone (33,143 reviews). \yun{The ground truth is defined based on the number of search results for each cost type on the user forum. We use this method to establish ground truth as an issue with more search results implies that the issue is encountered by more users, and thus collectively users care about it more.} Pearson correlation coefficient (PCC)~\citep{pearson} is utilized to evaluate the linear correlation between measured user concerns and \yun{the numbers of user views for the four performance cost types.}
The typical metric $\text{NDCG}@k$ (Normalized Discounted Cumulative Gain)~\citep{ndcg} is adopted for computing the accuracy in prioritizing issues, \yun{i.e., }

\begin{equation}
    \begin{split}
    \text{NDCG}@k & =\frac{\text{DCG}@k}{\text{IDCG}@k},\\
    \text{where} \; \text{DCG}@k & =\sum_{i=1}^k\frac{rank(i)}{\log_2(i+1)}, \\
    \text{and} \; \text{IDCG}@k &= \sum_{i=1}^{{rank}_k} \frac{rank(i)}{\log_2(i+1)}, 
    \end{split}
\end{equation}

\noindent where \add{$rank(i)$ indicates the ranking score of the $i$-th issue computed by Equ.~(2) and $rank_k$ in computing $\text{IDCG}@k$ represents the position list based on the computed ranking scores. The premise of $\text{DCG}$ is that highly important issues appearing lower in the prioritized results should be penalized as the ranking score is reduced logarithmically proportional to the position of the issues. IDCG (Ideal DCG) computes the maximum DCG based on the position list resulted by the ranking scores. $\text{NDCG}@k\in [0,1]$, and $k$ denoting the number of elements to be sorted. Higher $\text{NDCG}@k$ values represent that DCG values are close to the IDCG results, implying more accurate rankings.}
NDCG$@k$ is computed by comparing the rank of the measured user concerns for the four permanence cost types (e.g., CPU cost) with the rank of the user view numbers in the ground truth.

We measure users' concerns about the performance costs, including memory, CPU, network, and battery, of the Spotify Music apps based on RankMiner. Specifically, we capture top 50 (i.e., $k$=50) terms that are semantically close to the target cost type, e.g., battery. We further manually remove ambiguous and noisy ones from the captured top terms, such as the terms ``data volume'' and ``data plan'', shown in the box below. The remaining terms are battery-related terms. We also notice that misspelled words (\textit{e.g.}, ``batery'') can be captured through word embeddings. Table~\ref{tab:memory_issues} illustrates the terms related to each performance cost types.

\begin{center}
	\noindent\fbox{
		\parbox{0.45\textwidth}{
			\textbf{Battery-related terms:} \textit{battery life, \st{data volume}, \hl{batery}, battery power, \st{data plan}, battery juice, ...}
		}}
\end{center}

\begin{table}
	\small
	\center
	\caption{Performance-related terms.}
	\label{tab:memory_issues}
	\scalebox{0.8}{\begin{tabular}{|c||c|m{6.8cm}|}
		\hline
		Cost Type & \#Terms & \parbox[t]{6cm}{Related Term} \\
		\hline
		Memory & 19 & ram memory, storage, storage space, memory space, space, internal memory, ram, internal storage, internal space, disk space, gb, battery power, extra space, ram memory, unnecessary space, capacity, mb, valuable space, precious space \\
		\hline
		CPU & 17 & cpu, processor, gpu, cpu usage, laggy, slowly, too slow, incredibly slow, extremely slow, sluggish, painfully slow, terribly slow, take age, slower, slower than before, lag, fast \\
		\hline
		Network & 12 & network connection, data connection, wifi connection, network signal, wifi, wifi network, wifi signal, internet connectivity, wireless connection, 4g connection, internet connection, wireless network \\
		\hline
		Battery & 14 & battery life, battery power, batery, batt, battery drain, battery usage, battery rapidly, battery dry, battery overnight, battery juice, batterie, battery excessively, battery life, drain battery \\
		\hline
	\end{tabular}
}
\end{table}

\subsubsection{Results} Table~\ref{tab:ndcg} depicts the comparison results of our methods (sigmod and soft division methods) with two baseline methods, one is only based on the review percentage (i.e., the CrossMiner method~\cite{man2016experience}) and the other is based on the user sentiment ($R_{i}$ in Equation~\ref{eq:userconcern}). We validate the quantified users' complaints about the four types of costs (memory, CPU, network and battery). As Table~\ref{tab:ndcg} shows, our methods achieve the best properties than the basic methods in terms of both PCC and $\text{NDCG}@4$, where $\text{NDCG}@4$ measures the accuracy in ranking four types of costs. Specifically, the soft division and sigmod methods can better identify important performance issues, with an increase of 2.5\% for $\text{NDCG}@4$ compared to CrossMiner~\cite{man2016experience}. \add{For the PCC results, the soft division method surpasses the ratings-based method by 2.14 times in terms of the correlation coefficients, however the $p$-values ($>0.05$) show that the correlations are not statistically significant, which means the relations between different ranking scores and the ground truth may be weak. This could be attributed to the small sample size involved. Overall, the proposed methods can prioritize the issues more accurately by balancing review ratings and percentages.}
During the analysis, we adopt the soft division method, which achieves the most optimal results in our comparison, for scoring users' concerns.

\begin{table}
	\caption{Results of comparison with baselines. \add{The subscripts beside the correlation coefficients indicate the corresponding resulted $p$-values.}}
	\label{tab:ndcg}
	\center
	\scalebox{0.76}{%
	\begin{tabular}{|c||c|c|c|c|}
		\hline
		& \begin{tabular}{@{}c@{}}CrossMiner \\ (Percent-Based)\end{tabular} & Rating-Based & \begin{tabular}{@{}c@{}}RankMiner \\ (Sigmod)\end{tabular} & \begin{tabular}{@{}c@{}}RankMiner \\ (Soft Division)\end{tabular}\\
		\hline
		PCC & 0.728\textsubscript{0.272} & 0.253\textsubscript{0.747} & 0.783\textsubscript{0.218} & \textbf{0.794}\textsubscript{0.206} \\
		\hline
		NDCG@4 & 0.854 & 0.869 & \textbf{0.875} & \textbf{0.875} \\
		\hline
	\end{tabular}
}
\end{table}

\begin{center}
	\noindent\fbox{
		\parbox{0.45\textwidth}{
			\textbf{Finding 1:} \emph{The proposed RankMiner approach can effectively quantify user concerns about specific app issues.}
        }
    }
\end{center}

\subsection{RQ1: Do the performance costs of in-app ads significantly increase the no-ads versions?}
\subsubsection{Motivation}
We revisit some of the questions (i.e., what is the energy/network/memory/CPU cost of ads?) investigated by Gui \textit{et al.}~\cite{gui2015truth}. Differently  from previous work~\cite{gui2015truth}, which only uses one experimental usage trace per app, we collect practical usage traces of subject apps. We want to examine whether the performance costs of in-ads apps and their no-ads versions exist significant differences in a relatively more practical scenario.

\subsubsection{Methodology} 
\begin{figure}
    \centering
    \includegraphics[width=0.35\textwidth]{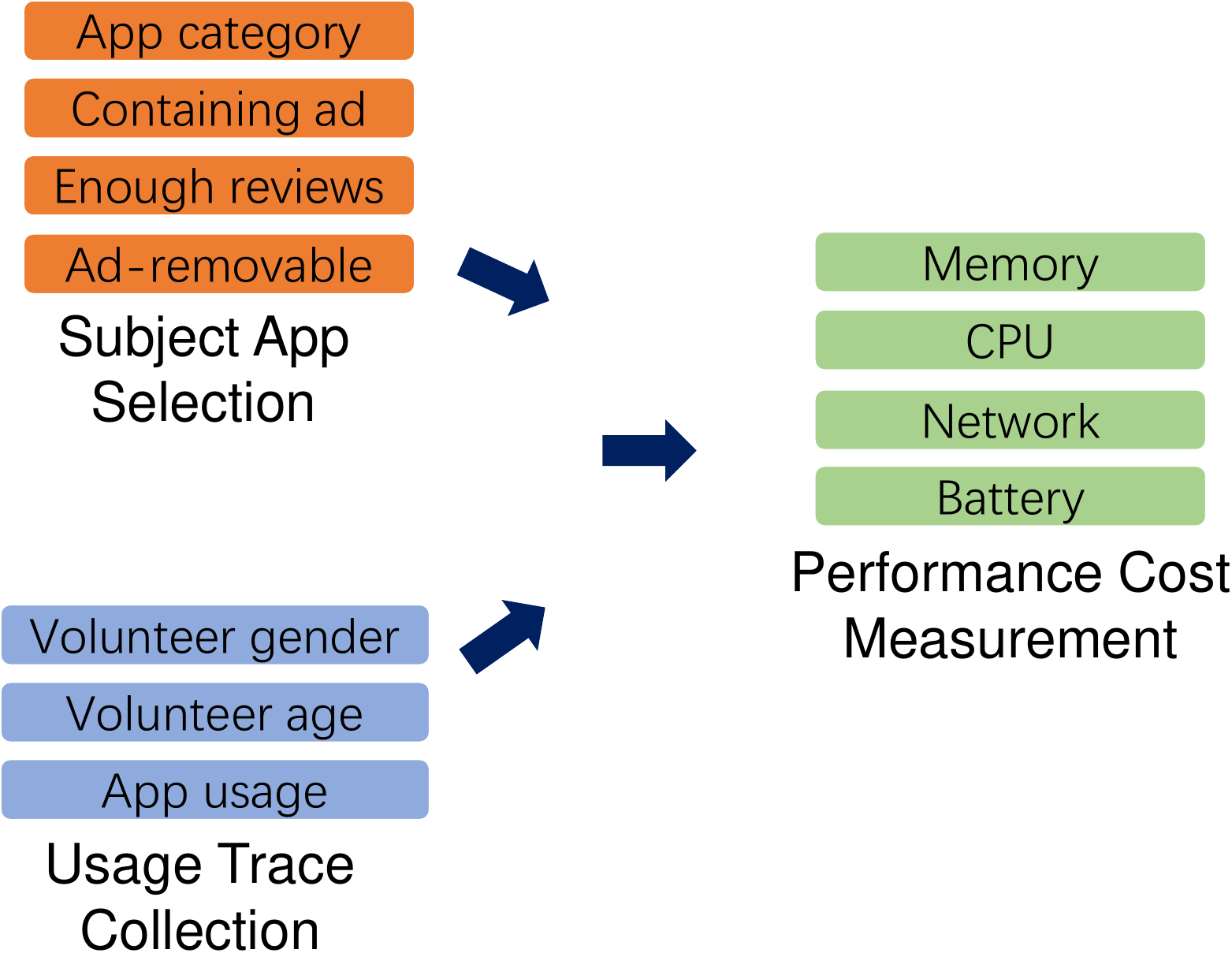}
    \caption{Workflow of performance costs of in-app ads.}
    \label{fig:ad_cost_framework}
\end{figure}

The workflow for measuring performance costs of in-app ads can be found in Figure~\ref{fig:ad_cost_framework}.

\begin{table*}
    \centering
    \caption{Subject apps used to answer RQ2 in our empirical study.}\label{tab:subjects}
	\scalebox{0.86}{
			\begin{tabular}{l|c|m{4.5cm}|l|m{1.3cm}|c|m{1cm}}
				\hline
				\hline
				Category & ID & App Name & Package Name & Version & \# Reviews & Overall Rating \\
				\thickhline
				\multirow{4}{*}{Weather} & A1 & RadarNow! & com.usnaviguide.radar\_now & 6.3 & 2,346 & 4.4 \\			
				& A2 & Transparent clock \& weather & com.droid27.transparentclockweather & 0.99.02.02 & 918 & 4.3 \\
				& A3 & Weather Underground: Forecasts & com.wunderground.android.weather & 5.6 & 4,584 & 4.5 \\
				& A4 & AccuWeather & com.accuweather.android & 4.6.0 & 8,691 & 4.3 \\
				\hline
				\multirow{4}{*}{Productivity} & A5 & QR \& Barcode Scanner & com.gamma.scan & 1.373 & 297 & 4.3 \\
				& A6 & Advanced Task Killer & com.rechild.advancedtaskkiller & 2.2.1B216 & 358 & 4.4 \\
				& A7 & Super-Bright LED Flashlight & com.surpax.ledflashlight.panel & 1.1.4 & 1,661 & 4.6 \\			
				& A8 & iTranslate - Free Translator & at.nk.tools.iTranslate & 3.5.8 & 242 & 4.4 \\			
				& A9 & AVG Cleaner for Android phones & com.avg.cleaner & 3.7.0.1 & 494 & 4.3 \\
				\hline
				\multirow{7}{*}{Health \& Fitness} & A10 & Pedometer & com.tayu.tau.pedometer & 5.19 & 2,024 & 4.4 \\
				& A11 & Pedometer \& Weight Loss Coach & cc.pacer.androidapp & 2.17.0 & 1,576 & 4.5 \\
				& A12 & Period Tracker & com.period.tracker.lite & 2.4.4 & 1,332 & 4.5 \\
				& A13 & Alarm Clock Plus\mystar  & com.vp.alarmClockPlusDock & 5.2 & 577 & 4.4 \\
				& A14 & Daily Ab Workout FREE & com.tinymission.dailyabworkoutfree1 & 5.01 & 25 & 4.4 \\
				& A15 & Map My Ride GPS Cycling Riding & com.mapmyride.android2 & 17.2.1 & 408 & 4.4 \\
				& A16 & Calorie Counter - MyFitnessPal & com.myfitnesspal.android & 6.5.6 & 2,267 & 4.6 \\
				\hline
				\multirow{4}{*}{News \& Magazines} & A17 & BBC News & bbc.mobile.news.ww & 4.0.0.80 & 9,693 & 4.3 \\
				& A18 & Fox News & com.foxnews.android & 2.5.0 & 4,163 & 4.5 \\
				& A19 & NYTimes - Latest News & com.nytimes.android & 6.09.1 & 71 & 3.8 \\
				& A20 & Dailyhunt (Newshunt) News & com.eterno & 8.3.17 & 1,452 & 4.3 \\
				\hline
				\hline		
			\end{tabular}
	}
\end{table*}

\textbf{Subject App Selection.} We select 20 popular apps from Google Play as the subjects based on the following four criteria: (1) they are selected from different categories - to ensure the generalization of our results; (2) they are apps containing ads; (3) they have a large number of reviews - indicating that user feedback can be sufficiently reflected in the reviews; and (4) they can be convertible to no-ads versions - for measuring the costs caused by ads. To collect apps that satisfy the first criterion, we randomly search the top 20 apps in each of the categories (except games and family apps) on Google Play. Since Google Play provides the number of reviews and declaration about ads, we extract apps with more than 10,000 reviews and with ads contained. To satisfy the last criterion, we convert these apps to no-ads versions based on {\tt Xposed} \citep{xposed} in a random order and then inspect whether the ads had been successfully removed. Finally, we choose 20 subjects for our experiment analysis. Their details are illustrated in Table~\ref{tab:subjects}, where we list the category, app name, package name, version, number of reviews, and overall rating for each subject app.

\textbf{Usage Trace Collection.} For rendering the viewing traces of ads various, 17 users are selected from different genders (six females and 11 males), and distributed in different age groups (six of them are aged at 18-25, ten at 25-30, and one at 30-35). All the selected participants satisfy the following criteria: 1) they interact with apps for more than 30 minutes daily - indicating that the users are familiar with using mobile apps; 2) they have experience using apps of different categories - considering the multi-categories of the subject apps; and 3) they are willing to spend time on our experiments - implying that they will take patience to execute the apps \textit{according to their usual habits}. We invite them to exercise the functionalities of the 20 subject apps according to their own usage habits. 

\yun{Table~\ref{tab:usagetrace} depicts the statistics of the duration for the collected user traces, including the maximum, minimum, and average duration for each app. We can observe that} the average interaction time for the apps ranges from 14 seconds to 2.48 minutes. Short interaction spans may be attributed to the simple functionality provided by some apps. For example, the app ``com.rechild.advancedtaskkiller'' \yun{(A6)} mainly supports service killing by clicking one button on the home page, which costs about 23 seconds on average according to our records. At least 70\% apps are executed for more than one minute on average, and only one app ``com.gamma.scan'' \yun{(A5)} is executed with less than 20 seconds.

\begin{table}
	\small
	\center
	\caption{\yun{Statistics of duration for collected usage traces.}}
	\label{tab:usagetrace}
	\scalebox{0.9}{\begin{tabular}{|c||r|r|r|}
	\hline
	ID & Max. (s) & Min. (s)& Avg. (s) \\
	\hline
		A1 & 155.12	& 66.36	& 19.89	\\
        A2 & 92.76 & 25.37 & 61.18 	\\
       A3 & 125.80 & 20.28 & 67.93	\\	
        A4&153.56 & 24.30 & 68.44 \\
        A5 &42.77& 0.07 & 14.51	\\
        A6	&	59.34 & 3.01 & 23.49 \\
        A7 &	69.36 &  4.48 & 23.50	\\	
    A8	& 167.59 & 28.37 & 65.30\\	
    A9	& 331.12& 56.34 &	13.24 \\
    A10	& 143.03 & 11.34 & 58.35 \\		
    A11	& 153.71 &  11.34 &  64.44	\\	
    A12	& 154.18 & 33.79 & 96.72\\
    A13	&134.44 & 20.98 & 69.79	\\	
    A14	& 210.15 & 25.74 & 105.03	\\	
    A15	& 230.95 & 22.46 & 102.43	\\	
    A16	& 501.06 & 13.57& 149.52	\\
    A17	 & 325.52 & 15.10 & 96.64	\\	
    A18	& 292.97 & 6.88 & 88.64	\\
    A19	& 243.32 & 27.35 & 100.29\\	
    A20	& 190.19 & 11.92 & 87.72\\
\hline
	\end{tabular}
}
\end{table}

For each app, we measure 102 times\footnote{102 = 17$\times$6, where 17 is the number of volunteer users and six denotes the total measuring times for both the with-ads and no-ads versions of an app.} by repeating both the with-ads version and the no-ads version three times using {\tt RERAN}~\citep{gomez2013reran}. \add{Whether the differences of the collected statistics for the 102 runs on each app are significant or not is not examined.} The average values are calculated to alleviate noises. To mitigate background noise, we restore the system environment to its original state before each version execution. Then we install the app and start its execution. When a subject app is launched, the tools {\tt tcpdump} and {\tt top} are started to capture the transmitted data traffic and memory/CPU consumption, respectively. We also monitor the app execution to ensure that they are consistent with the records. Note that even though running tools, such as {\tt top} and {\tt Xposed}, can affect mobile performance, the effects could be consistent on both versions (with-ads and no-ads)~\citep{gui2015truth} and can thus be ignored in our cost measurement. Overall, we measure the 20 subject apps 2,040 times\footnote{2040 = 102$\times$20, where 20 denotes the number of subject apps.} in total. The whole measurement process lasted for for more than one month.

\textbf{Performance Cost Measurement.}
We measure the memory, CPU and network costs following~\citet{gui2015truth}, and battery cost following~\citet{DBLP:conf/kbse/GaoZSLK18}. The ad costs are computed by subtracting the costs of no-ads versions from those of with-ads versions.

\subsubsection{Results}
\begin{figure*}
	\centering
	\begin{tabular}{cc}
	\includegraphics[width=0.36 \textwidth]{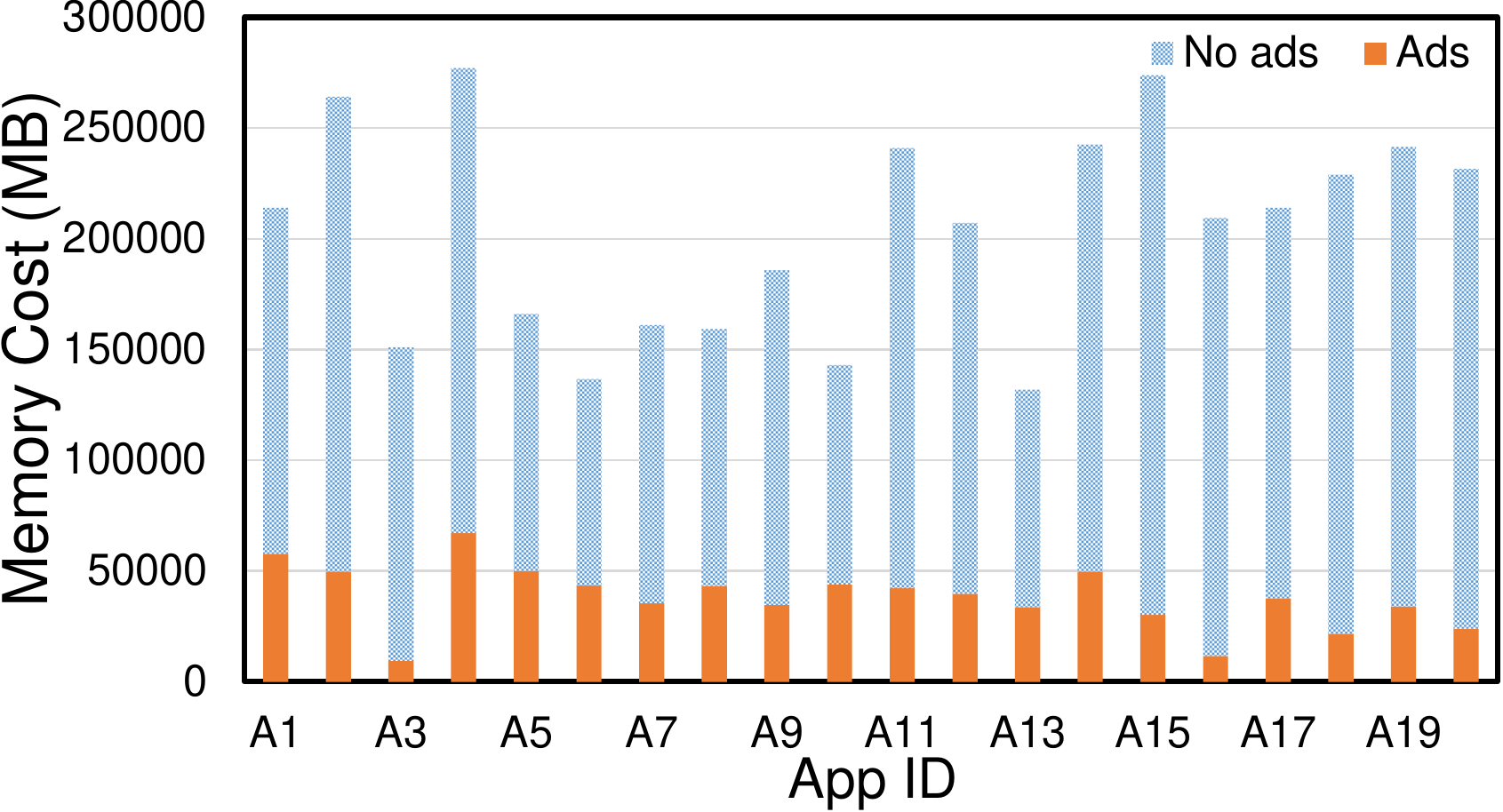} & \includegraphics[width=0.32 \textwidth]{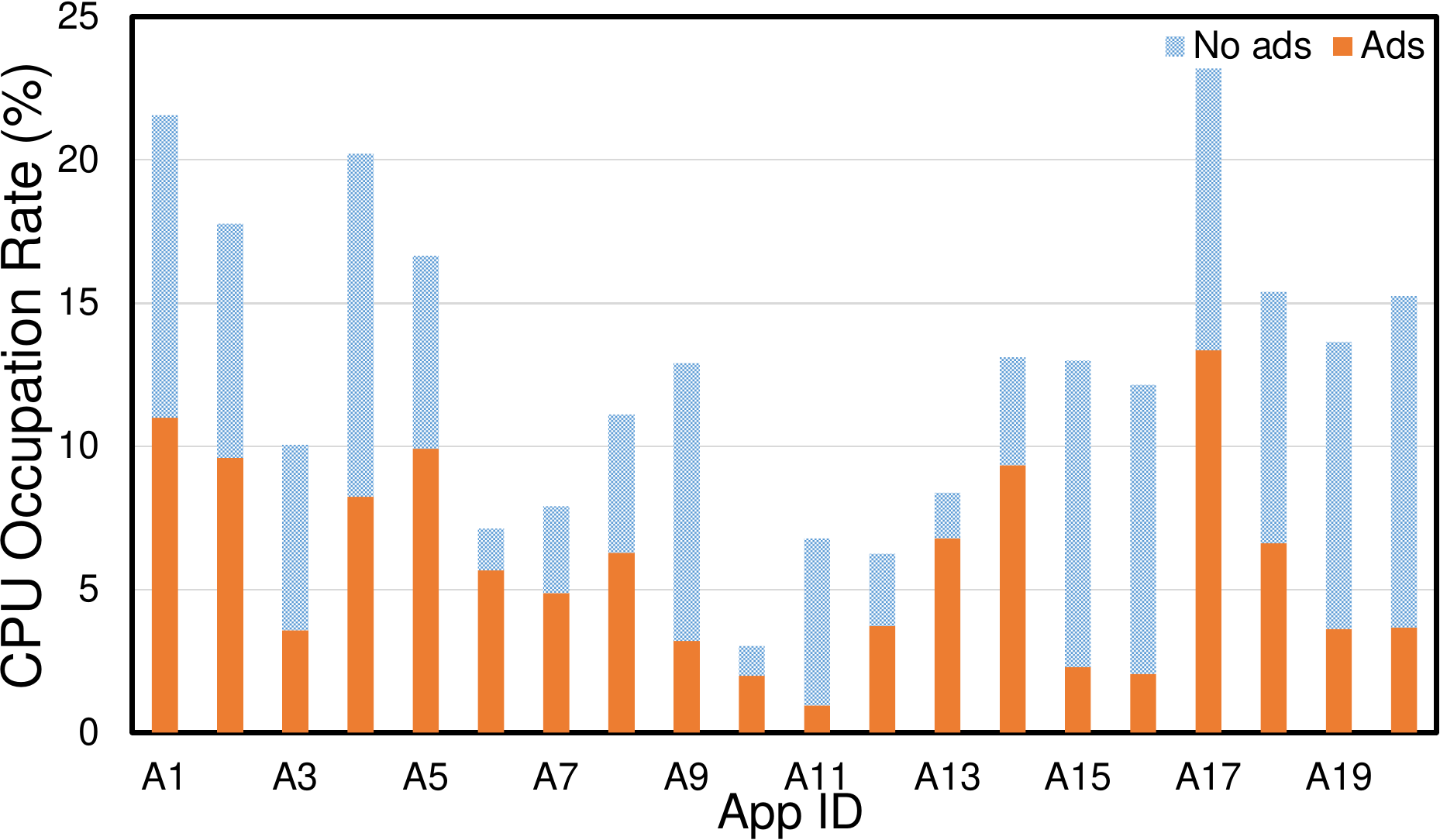} \\
	 [-0.02in]
	\small{(a) Memory} & \small{(b) CPU}\\
	\includegraphics[width=0.33 \textwidth]{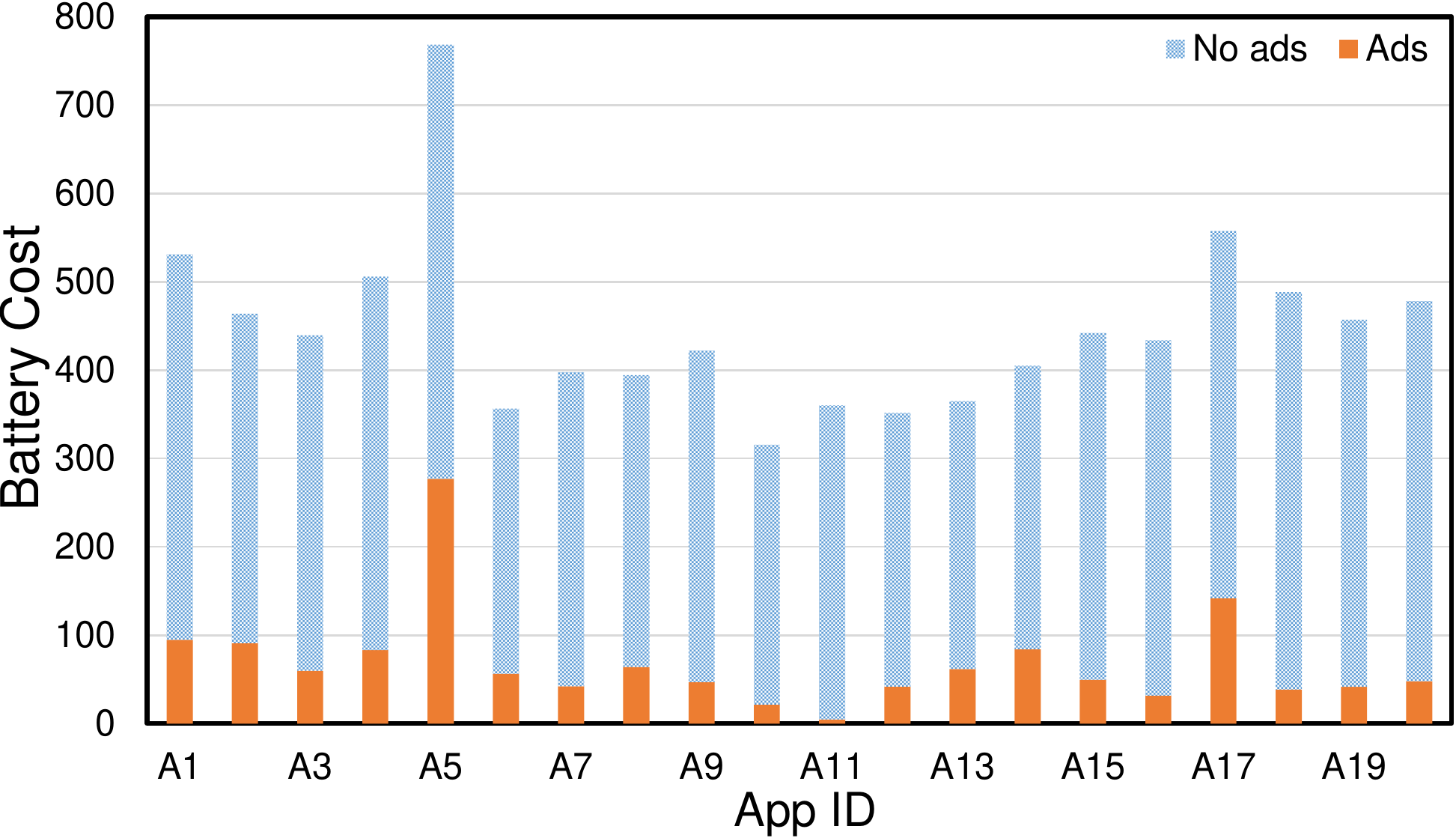} & \includegraphics[width=0.35 \textwidth]{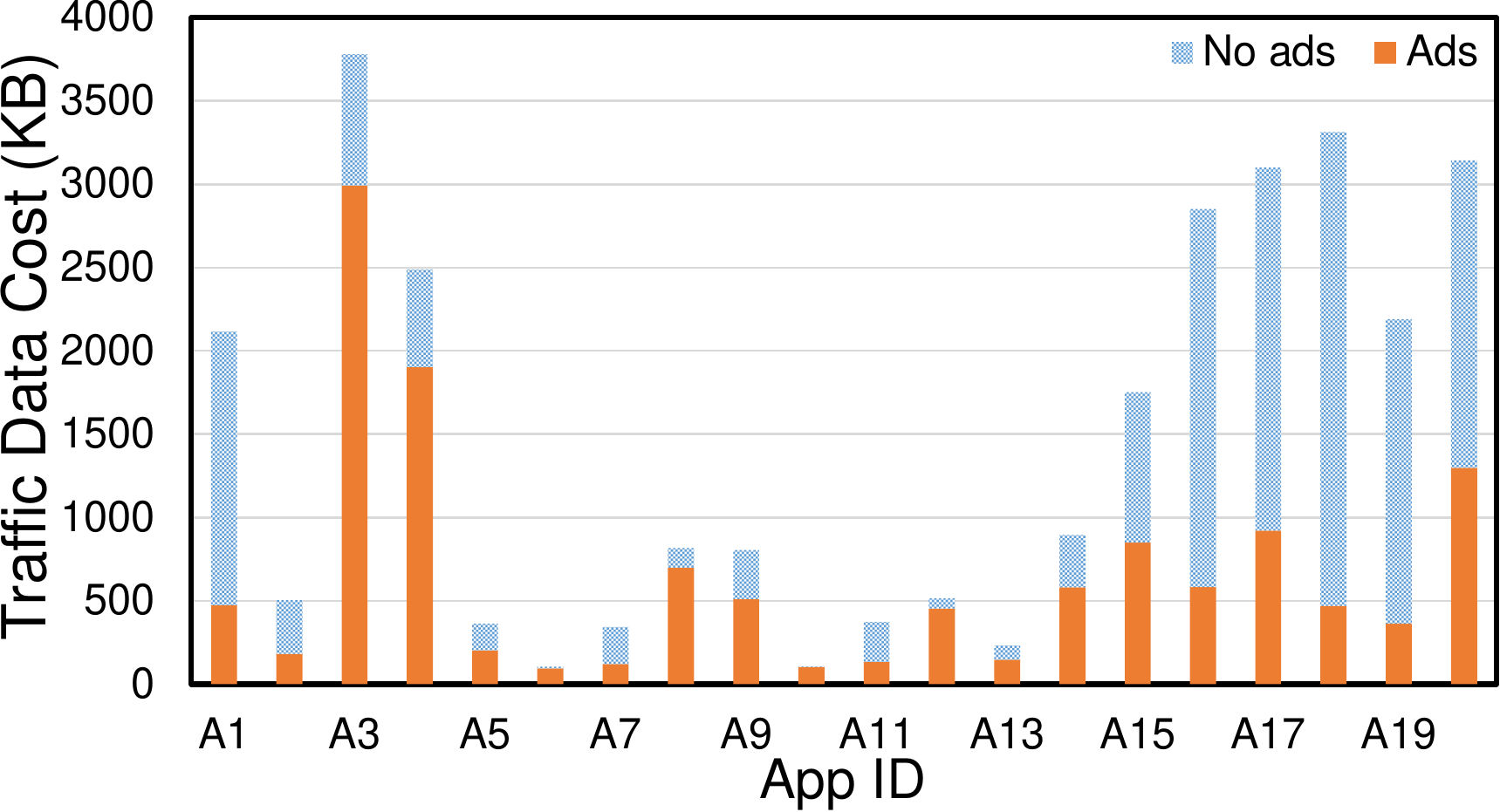}\\
	 [-0.02in]
	\small{(c) Battery} & \small{(d) Traffic} \\
	\end{tabular}
	\caption{RQ3: Performance consumption of with-ads (in orange) and no-ads versions (in light blue).}
	\label{fig:memcost}
\end{figure*}

For each subject app, we measure the four types of performance costs (\textit{i.e.}, memory, CPU, network and battery consumption) for both with-ads and no-ads versions. Figure~\ref{fig:memcost} depicts the costs of the 20 apps, with blue bars indicating the memory costs of the no-ads versions and orange bars representing the ad costs. According to the figure, all the with-ads versions consume more performance cost than their no-ads versions. \yun{For example, with ads integrated, the CPU cost of A10 has apparently increased, and the network usage of subjects such as A6 and A12 shows dramatic growth.} The memory increase ranges from 5.9\% (A16) to 46.4\% (A6), with an average of 25.2\%. For CPU cost, ads in the subject apps consume 1.0\% to 12.0\% with respect to the CPU occupation rate, with median cost at 7.4\%. This indicates that mobile ads indeed influence the device resource, which is consistent with the results by Gui \textit{et al.}~\citep{gui2015truth}.

Table~\ref{tab:costresult} shows the statistics of all measured performance costs for the 20 subjects, with the average increase rate and corresponding deviation (which represents the cost increase variations among the subject apps). Network usage has the most remarkable increase (113.9\%) on average. The distinct cost increase (s.d. at 108.9\%) of network usage may be attributed to the ads-oriented design of some apps. CPU costs experience a modest increase (6.9\% on average). Moreover, the growth in battery drainage is also noteworthy, with the average increase at 17.7\% and deviation at 11.9\%. Heavy performance costs may ruin user experience and drive users to uninstall the apps, which is the reason why developers and researchers pay attention to ad performance costs. 


\begin{table}
	\small
	\center
	\caption{Average and standard deviation of the increase rate of performance cost when comparing with-ads version with the no-ads version.}
	\label{tab:costresult}
	\scalebox{0.9}{\begin{tabular}{|m{3cm}||c|c|c|c|}
		\hline
		Cost Type & \parbox[t]{1.2cm}{Memory} & \parbox[t]{0.7cm}{CPU} & \parbox[t]{1.0cm}{Network} & \parbox[t]{0.8cm}{Battery} \\
		\hline
		Average & 25.2\% & 6.9\% & \textbf{113.9\%} & 17.7\% \\
		\hline
		Standard Deviation & 12.5\% & 3.7\% & \textbf{108.9\%} & 11.9\% \\
		\hline
	\end{tabular}
}
\end{table}

We further observe whether statistically significant differences exist between performance costs of with-ads versions and those of no-ads versions. We first check the distributions of each type of measured performance costs by the Shapiro-Wilk test~\citep{shapiro1965analysis}. The Shapiro-Wilk test is a typical test of normality of which the null hypothesis is that the input samples come from a normally distributed population. If the p-value computed by the Shapiro-Wilk test is smaller than 0.05, we achieve that the input distribution is significantly different from normal distribution. Table~\ref{tab:normality} lists the p-value results of Shapiro-Wilk test for different performance cost types. We observe that except for the traffic cost of with-ads versions, all the other measured costs render normal distributions. This may be  because traffic is more sensitive to the usage pattern and time of various users.
Therefore, for memory, CPU and battery costs, we use the \textit{paired t-test}~\citep{hsu2008paired} for comparing the distributions between with-ads and no-ads versions, and use the Wilcoxon signed-rank test for analyzing the traffic costs. The paired t-test is a statistical test to determine whether the mean difference between paired observations is zero, with the p-value less than 0.05 indicating the difference between two paired inputs is significant. We use paired t-test for costs of memory, CPU, and Battery, because the subject apps may have different cost values for with-ads and no-ads versions, and the differences between pairs are normally distributed. The Wilcoxon signed-rank test is a paired version of the Wilcoxon rank-sum test. 

\begin{figure*}
	\centering

    \begin{subfigure}{.24\textwidth}
    \centering
    \includegraphics[width=0.98 \textwidth]{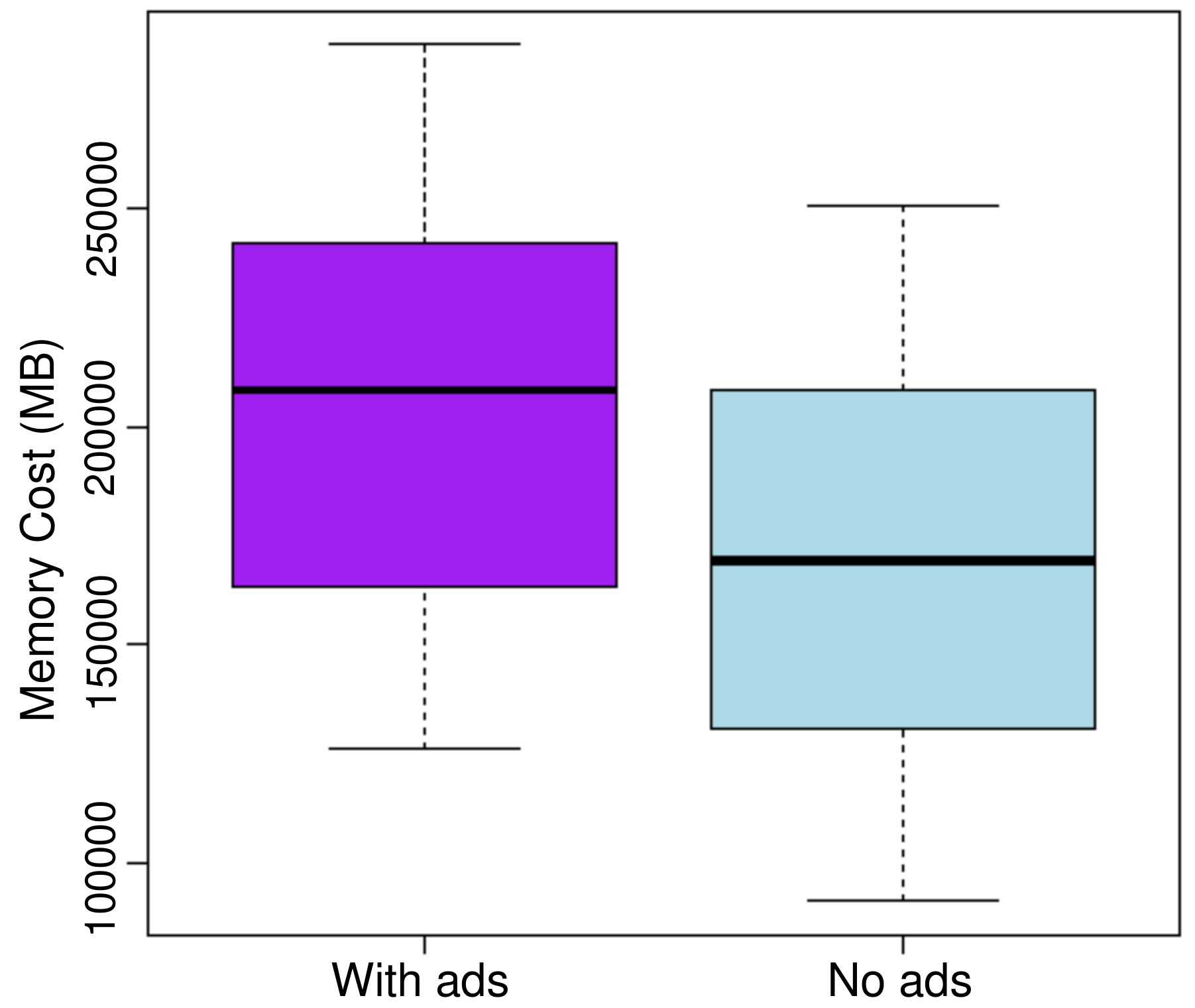}
    \caption{\small{Memory Cost}}
    \end{subfigure}
    \begin{subfigure}{.25\textwidth}
    \centering
    \includegraphics[width=0.99 \textwidth]{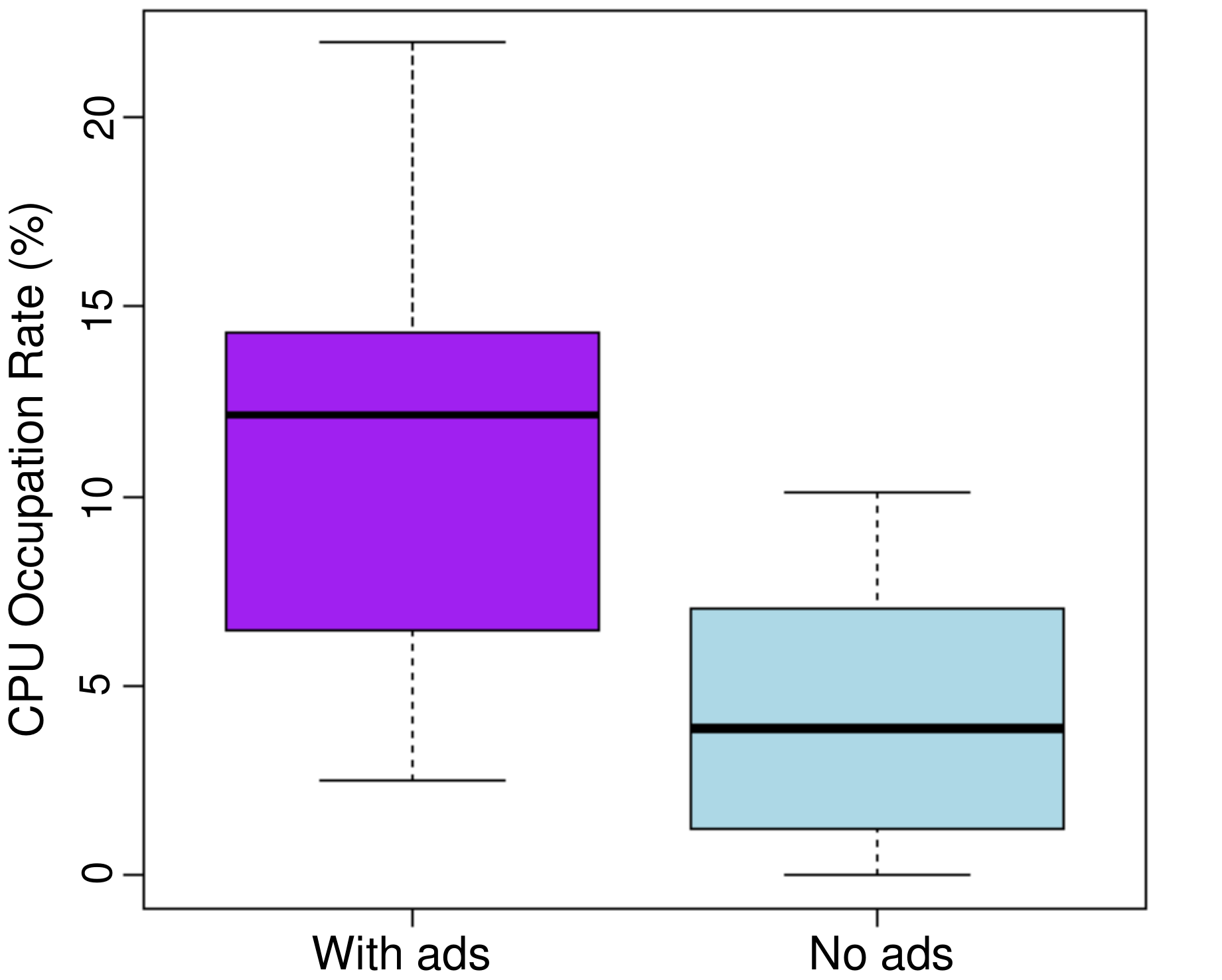}
    \caption{\small{CPU Cost}}
    \end{subfigure}
    \begin{subfigure}{.24\textwidth}
    \centering
    \includegraphics[width=0.98 \textwidth]{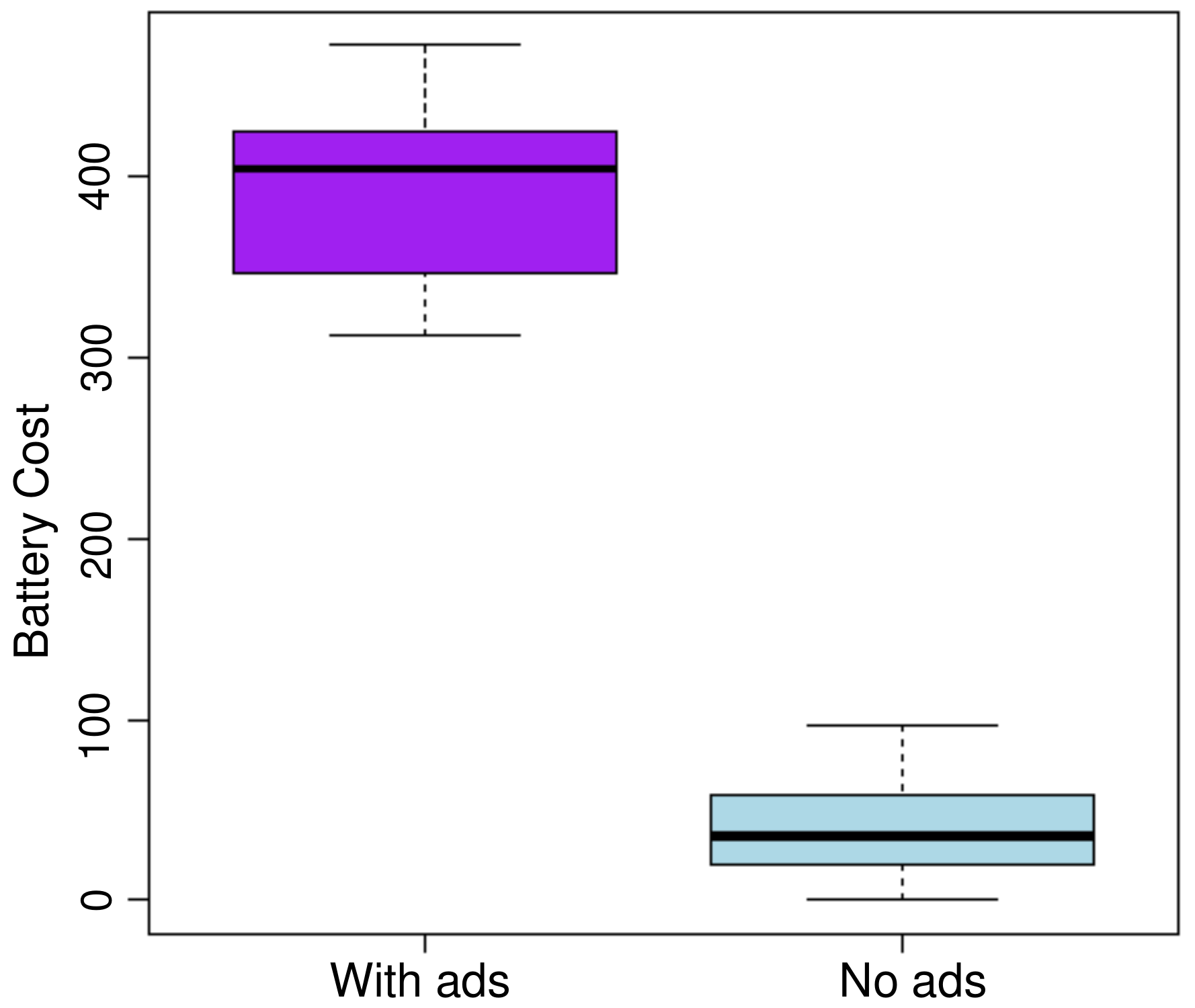}
    \caption{\small{Battery Cost}}
    \end{subfigure}
    \begin{subfigure}{.24\textwidth}
    \centering
    \includegraphics[width=0.98 \textwidth]{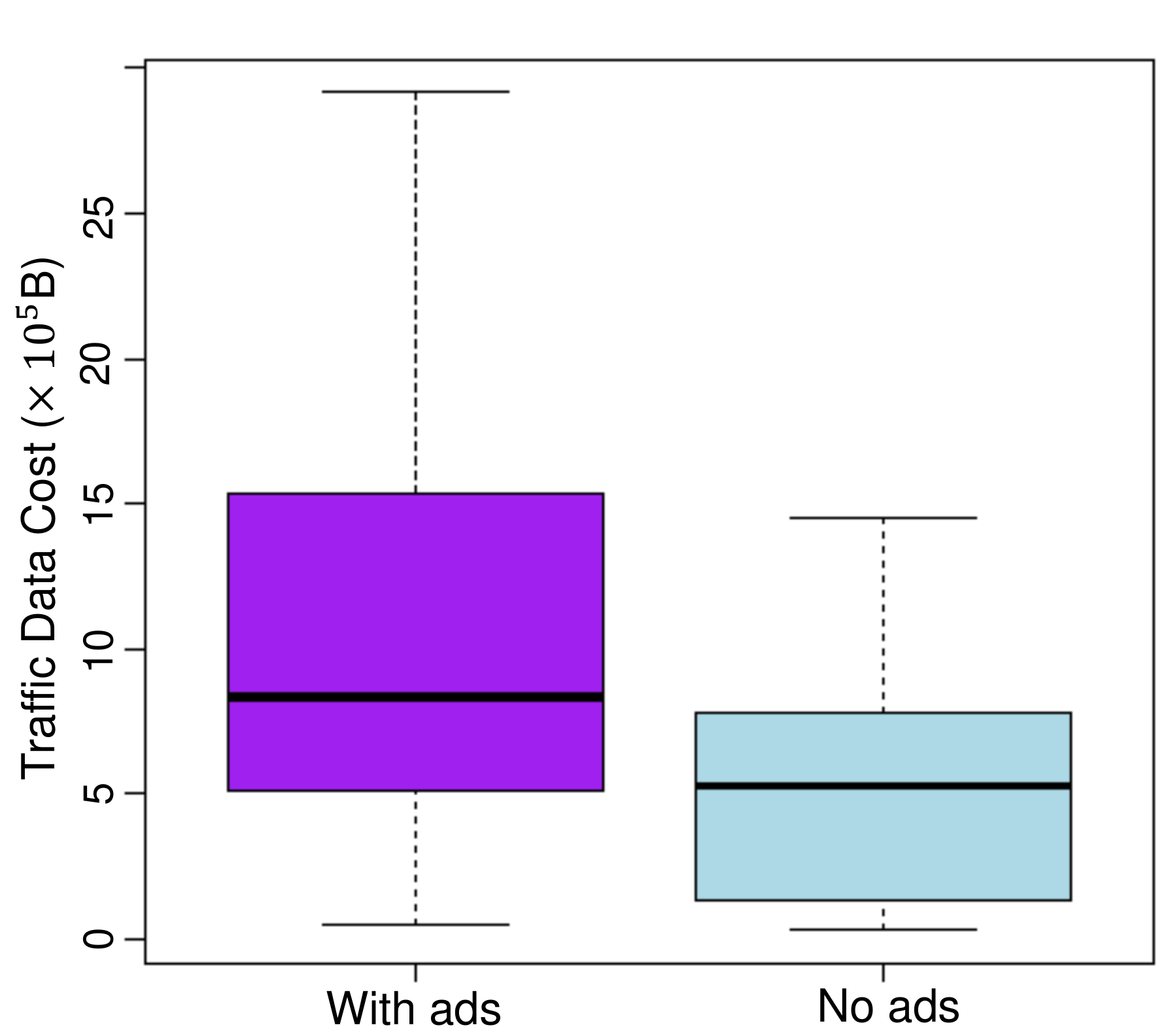}
    \caption{\small{Traffic Cost}}
    \end{subfigure}
	\caption{Performance cost distributions for with-ads (in purple) and no-ads versions (in light blue).}
	
	\label{fig:sigcost}
\end{figure*}

Figure~\ref{fig:sigcost} illustrates the comparison on the performance costs of with-ads and no-ads versions. The p-values in paired t-test and Wilcoxon signed-rank tests show that the two input distributions are significantly different. The effect sizes measured by Vargha and Delaney's
$A_{12}$ are all negligible. The results indicate that versions with ads expend significantly more performance costs, which is consistent with the studies in \citep{gui2015truth} and \citep{DBLP:conf/iwpc/SaboridoKAG17}.

\begin{center}
	\noindent\fbox{
		\parbox{0.45\textwidth}{
			\textbf{Finding 2:} \emph{Performance costs of with-ads versions are significantly larger than those of no-ads versions.
			}
			}}
\end{center}

\begin{table}
	\small
	\center
	\caption{Normality test of differences between measured performance costs of with-ads versions and no-ads versions. The p-value$<$0.05 means the differences are not normally distributed. }
	\label{tab:normality}
	\scalebox{0.9}{\begin{tabular}{|l||r|r|r|r|}
		\hline
		Cost Type & Memory & CPU & Battery & Traffic \\
		\hline
		p-value & 0.666 & 0.116 & 0.429 & \textbf{0.001}\\
		\hline
	\end{tabular}
}
\end{table}

\subsection{RQ2: How can the performance costs of ads affect user opinions?}
\subsubsection{Motivation} We aim at exploring whether users show more concerns for more performance costs of in-app ads, and which performance cost users care more about. Thus, developers can understand more about user perceptions of in-app ads, and pay more attention to user-concerned performance costs.

\subsubsection{Methodology}  We crawl totally 34,455 reviews published from December, 2016 to April, 2017 for the 20 apps. The reviews are large enough for review analysis~\citep{chen2014ar}, which can effectively capture the user experience. To ensure that user reviews are specific to subject app versions, we select the reviews posted by users within two months\footnote{The period is defined following previous work~\citep{DBLP:conf/wcre/CiurumeleaSPG17}.} after the corresponding version release.

We first retrieve \textbf{ad-related reviews} by extracting the reviews explicitly related to ads, \textit{i.e.}, reviews containing words such as ``ad'', ``ads'', or ``advert*'' (with regular expression)~\cite{gui2015truth}. Then we measure users' concerns about the performance costs, including memory, CPU, network, and battery, of both the with-ads apps and in-app ads based on RankMiner. We calculate user concerns about ads' performance costs based on the ad-related reviews only.

\subsubsection{Results} 
\begin{figure*}
	\centering
	\begin{tabular}{cc}
	\includegraphics[width=0.34 \textwidth]{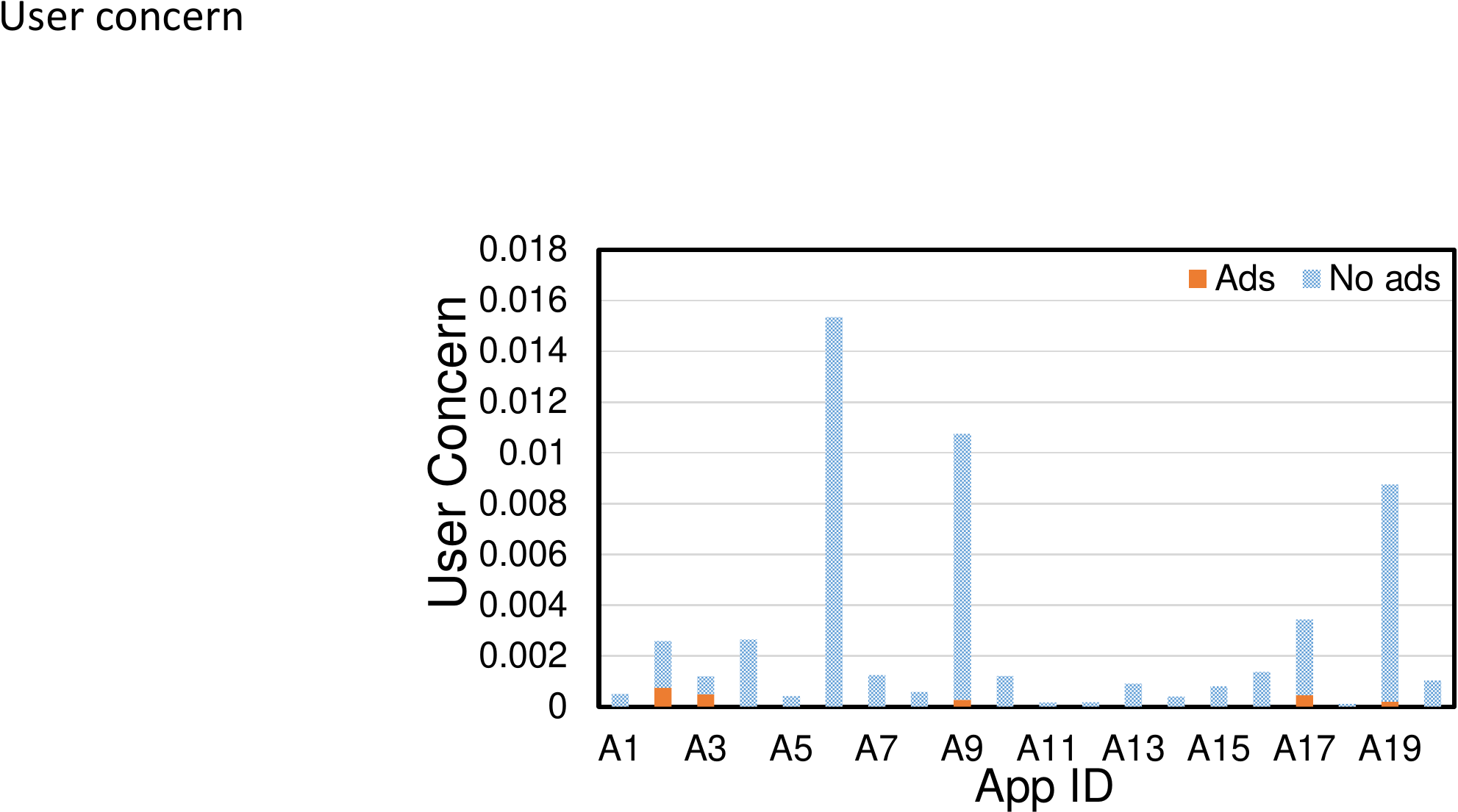} & \includegraphics[width=0.35 \textwidth]{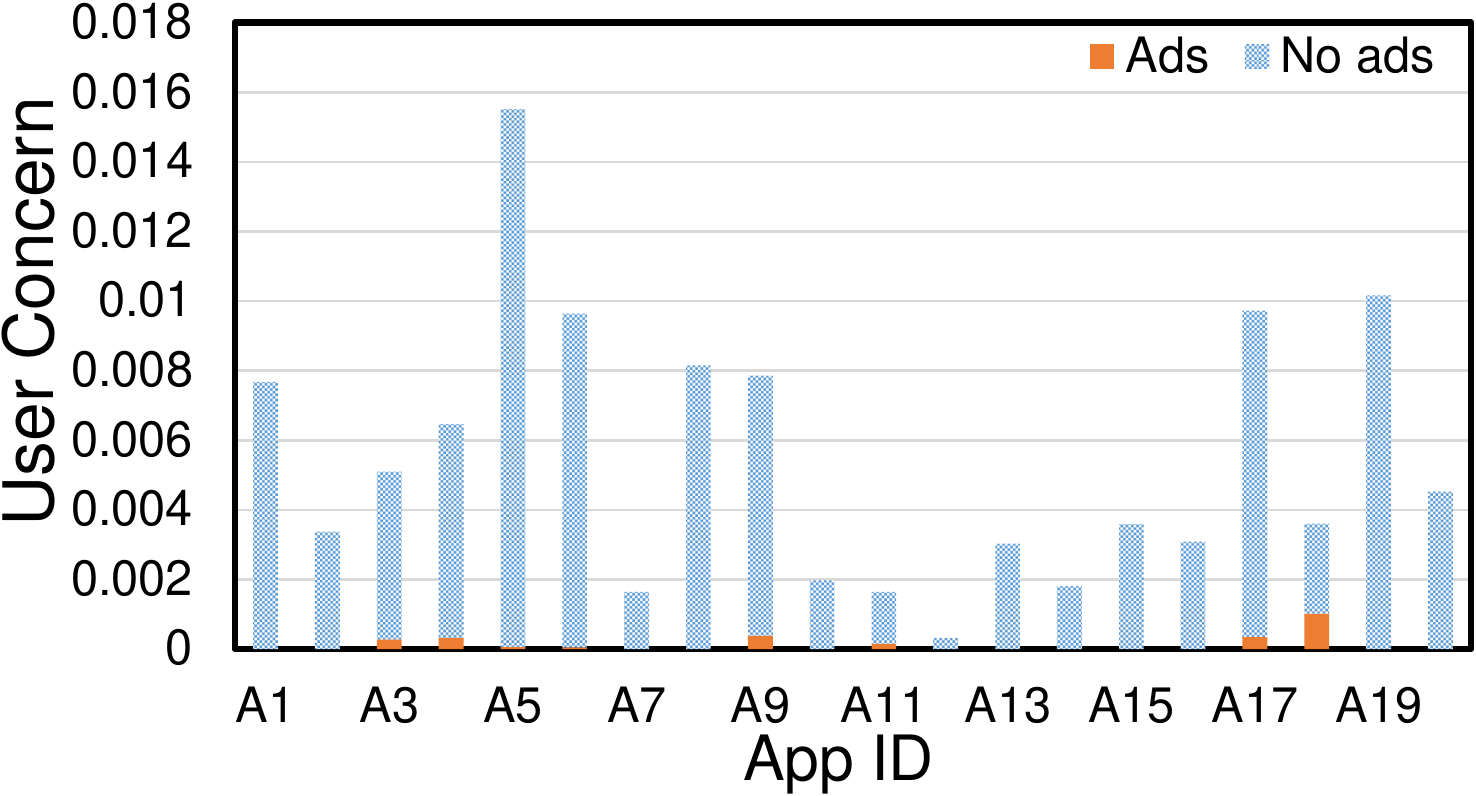} \\
	\small{(a) Memory Cost} & \small{(b) CPU Cost}\\
	\includegraphics[width=0.35 \textwidth]{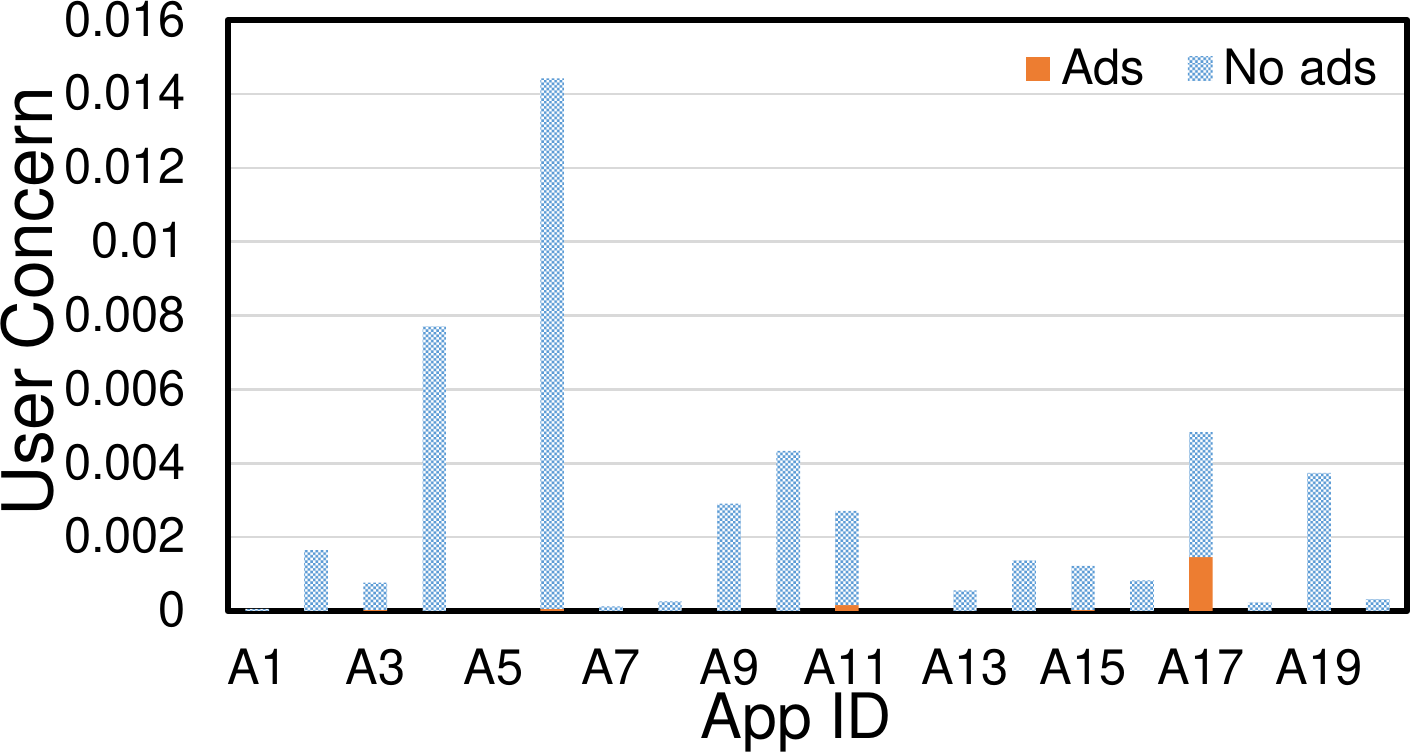} & \includegraphics[width=0.34 \textwidth]{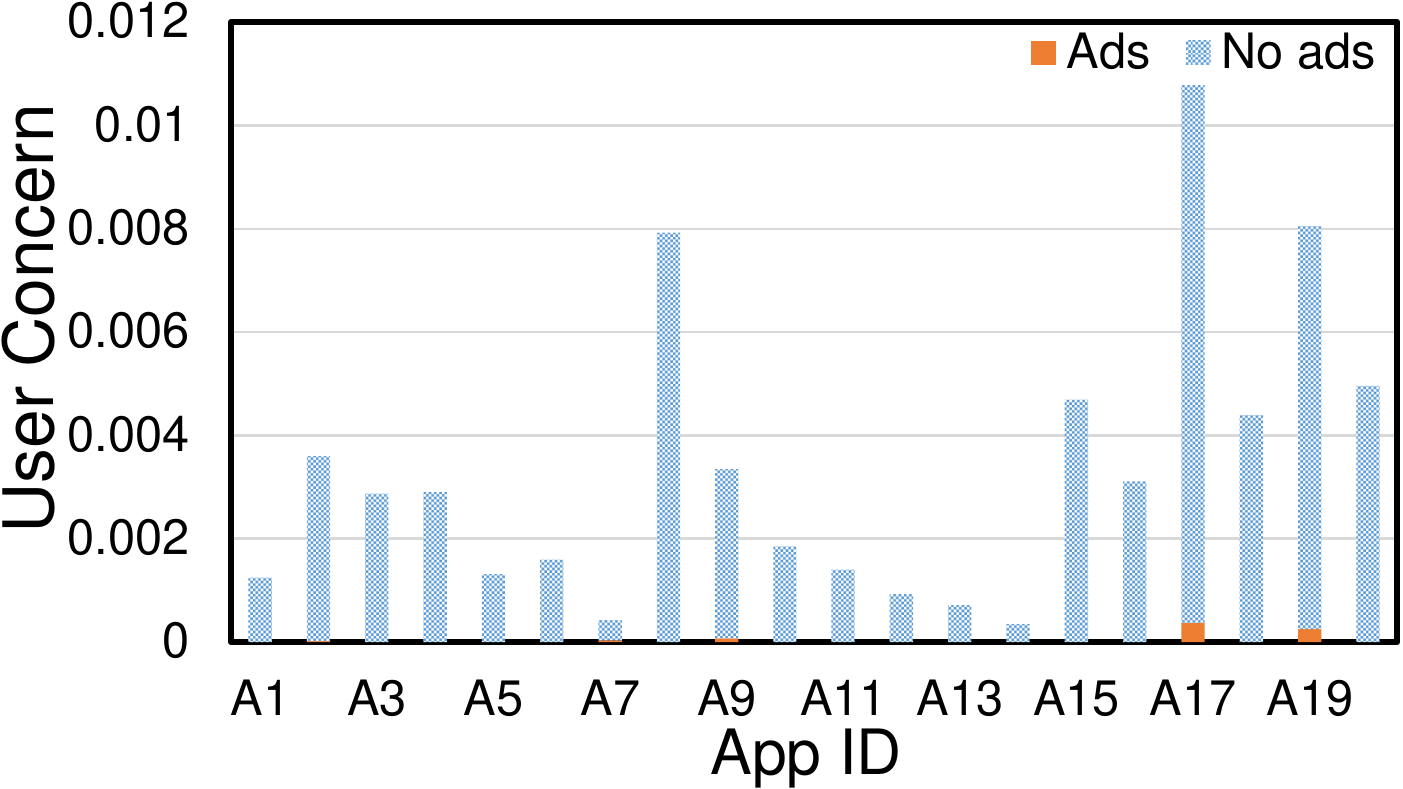}\\
	\small{(c) Battery Cost} & \small{(d) Traffic Cost} \\
	\end{tabular}
	\caption{Quantified user concerns about different performance cost types of the 20 subject apps.}
	\label{fig:memuser}
\end{figure*}

We illustrate the results of users' concerns about the performance costs in Figure~\ref{fig:memuser}, with the blue bars and orange bars denoting the measured values for no-ads and with-ads versions respectively. For the 20 subjects, users express different levels of concerns about the memory overhead of the in-app ads. For example, for the memory cost, A2 receives the most complaints about ads among all the subject apps, with an obvious increase of 35.9\% compared with the no-ads version \yun{indicated by blue bar in Figure~\ref{fig:memuser}}. By inspecting A2, we discover that in-app ads can occupy almost the whole screen space, especially with one banner on the top and one rectangle ad appearing in the middle when sliding downward. Interestingly, we find that 15 (75\%) apps receive zero negative feedback about the memory costs of ads \yun{(i.e., only blue bar is shown for the app in Figure~\ref{fig:memuser})}, such as A1. This implies that in most cases, user tend to be insensitive to the memory costs caused by in-app ads.

By observing the increase rate of quantified user concerns about all performance costs (shown in Table~\ref{tab:usercost}), we identify that memory costs have the largest rate of growth in user concerns (6.3\% on average) and the most obvious deviation (17.0\%) among the 20 apps. However, users express the least concerns about network costs, with the increase rate averaging at 0.9\% and a deviation of 1.9\%. Such an observation is different from what we have discovered in Table~\ref{tab:costresult}, where network costs exhibit the highest increase among all the performance costs. We find that 15/20, 12/20, 15/20, and 15/20 of the subject apps do not receive any complaints from users regarding the cost of memory, CPU, battery, and traffic, respectively. We think that users may perceive different types of performance costs differently. We then conduct correlation analysis to explore that there are strong correlations between user concerns and performance costs of ads. \yun{Specifically, we use PCC to calculate the correlations between the quantified user concerns and measured costs of the 20 subjects for each performance cost type.}



\begin{table}
	\caption{Increase rate of quantified user concerns about performance costs.}
	\label{tab:usercost}
	\center
	\scalebox{0.7}{\begin{tabular}{|m{3cm}||r|r|r|r|}
			\hline
			Cost Type & \parbox[t]{1.5cm}{Memory} & \parbox[t]{1.5cm}{CPU} & \parbox[t]{1.5cm}{Network} & \parbox[t]{1.5cm}{Battery} \\
			\hline
			Average & \textbf{6.3\%} & 3.5\% & 0.9\% & 2.7\% \\
			\hline
			Standard Deviation & \textbf{17.0\%} & 8.9\% & 1.9\% & 9.6\% \\
			\hline
		\end{tabular}
}
\end{table}

The correlations between performance costs and the corresponding user concerns are illustrated in Table~\ref{tab:correlation}. Almost all the PCC results indicate that their linear correlations are weak, especially for memory usage which represents nearly no correlation with the quantified user concerns (with PCC score $r_p=-0.132$). The only one performance type that presents moderate correlation with the quantified user concern is battery cost, with $r_p=0.534$ and $p=0.015<0.05$.

The results of PCC are consistent with those of SRC, where user concern shows a strongly increasing trend with more battery consumed ($p=0.0009\ll0.05$). This allows us to conclude that users care most about the battery cost among all the performance cost types. We attribute this to that the consumption of battery is more sensible than other cost to users, and therefore more battery costs tend to cause more unfavorable reviews.

We also observe the negative correlation between network cost and the corresponding user concern with respect to both PCC and SRC analysis. This means that more network costs could possibly bring better user experience. This might be against our common sense. We attribute this to the ubiquity of WiFi leading to fewer concerns about traffic consumed. According to~\citep{morewifi}, over 90\% of users choose WiFi connections when using smartphones. We therefore conclude that the network consumption of ads may not be concerned to users.

For CPU costs, the PCC ($r_p=0.166$) and SRC ($r_s=0.213$) scores display weak correlations with user concerns. The result is predictable, as users may not perceive the CPU cost on their mobile phones, and would generally think the crash or laggy performance is caused by mobile systems or app-specific functionalities. We conclude that the effect of CPU consumption on users is weak. Note that since our data are not time-series, causal impact analysis~\citep{martin2016causal,brodersen2015inferring} is not applicable in our situation. Moreover, our correlation analysis is applicable and convincing to determine the correlations between the two factors.

\begin{table}
	\caption{Correlation test results between performance costs of ads and user concerns.}
	\label{tab:correlation}
	\center
	\scalebox{0.62}{
		\begin{threeparttable}
			\begin{tabular}{|c||*2{>{\centering\arraybackslash}m{1cm}}|*2{>{\centering\arraybackslash}m{1cm}}|*2{>{\centering\arraybackslash}m{1cm}}|*2{>{\centering\arraybackslash}m{1cm}}|}
				\hline
				\multirow{2}{*}{Cost Type} & \multicolumn{2}{c|}{Memory} & \multicolumn{2}{c|}{CPU} & \multicolumn{2}{c|}{Network}  & \multicolumn{2}{c|}{Battery} \\
				\cline{2-9}
				 & $r$-score & $p$-value\tnote{3} &  $r$-score & $p$-value &  $r$-score & $p$-value &  $r$-score & $p$-value  \\
				\hline
				PCC\tnote{1} & 0.132 & 0.578 & 0.166 & 0.482 & -0.281 & 0.229 & 0.534 & \textbf{0.015}\\
				SRC\tnote{2} & 0.372 & 0.105 & 0.213 & 0.366 & -0.127 & 0.591 & 0.679 & \textbf{0.0009} \\
				\hline
			\end{tabular}
			\begin{tablenotes}
				\item[1,2] The absolute values of the PCC/SRC scores $r$ represent very weak correlations if $|r|<0.2$, weak correlation if $0.2\leq |r|<0.4$, moderate correlations if $0.4\leq |r|< 0.6$, strong correlations if $0.6\leq |r|<0.8$, and very strong correlation if $|r|\geq0.8$~\citep{evans1996straightforward}.
				\item[3] $p<0.05$ indicates that the correlation is statistically significant.
			\end{tablenotes}
		\end{threeparttable}
	}
\end{table}

\begin{center}
	\noindent\fbox{
		\parbox{0.45\textwidth}{
			\textbf{Finding 3:} \emph{Users care most about the battery cost among all the performance cost types, and show least sensitivity to the data traffic cost of ads. 
			}
			}}
\end{center}
\section{Discussion and Limitation}\label{sec:limitation}
In this section, we discuss the threats to the validity of our study and illustrate the steps we have taken to mitigate them. \yun{We also discuss the usefulness of our findings.}

\subsection{\yun{Threats to validity}}
\textbf{External Validity:} First, our experimental study is based on limited real apps from Google Play. Although the study does not involve
other app distribution platforms (\textit{e.g.}, App Store), we believe our results would also work across the board, since ad rendering mechanisms are similar across app markets. \yun{We determine the number of subjects following prior work~\cite{gui2015truth}, which achieves the finding that apps with ads have more hidden cost than those without ads based on 20 subject apps. In this paper,} we alleviate this threat by ensuring that the subject apps are popular apps distributed in four different categories. \add{We argue that future replications with similar contexts, e.g., using similar apps created in similar organizations, are likely to achieve identical observations as ours.} Future work will consider more apps and app platforms. Second, we collect usage traces from 17 volunteers which account for a limited number of the whole audience. In our experiments, ad displaying periods can impact the measured performance costs of ads. Since ads are generally set to refresh about every 60 seconds~\citep{refreshrate}, the collected usage traces would cover different situations of ad rendering and reloads (with the minimum interaction spans range from 0.06s to 33.8s and the maximum from 42.8s to 8.4min for the apps). Moreover, we invite the volunteers from different age groups and genders, which enriches the usage traces of in-app ads. Besides, there are no available datasets about the performance costs of ads or systematic tools for collecting all the performance consumed by ads on a large scale. Our work is the first to explore the performance costs of ads in practice.


\textbf{Internal Validity:} First, we leverage the {\tt Xposed} and {\tt AdBlocker Reborn} modules for generating no-ads versions, which may introduce additional workload to mobile apps. Since we instrument {\tt Xposed}, which has been widely used in performance testing and bug detection~\citep{kang2016persisdroid,casati2017exploiting}, into the phones for both with-ads and no-ads versions, the influence of {\tt Xposed} is consistent and can be eliminated by subtracting the costs of the two versions. We then just verify the influence of {\tt AdBlocker Reborn} on mobile performance. The costs are measured for three apps (including MediCalc, Google Maps, and RealCalc Plus) with the module enabled and disabled, respectively. The results exhibit that the average increase rates in costs are 3.0\%, 0.6\%, and 0.0\% for the memory, CPU, and battery, respectively. Compared with the performance consumption of each subject app, such cost increase is negligible.


Second, user concerns are quantified based on the proposed RankMiner and user reviews, which might not represent the real opinions of some users. To verify the effectiveness of RankMiner, we compare with baselines and our soft-division based method shows significant increase in accuracy (\textit{e.g.}, 214\% increase compared to the rating-based method in PCC). Besides, Google Play does not provide access to all the user reviews. Hence, any analysis on the user reviews might encounter dealing with an incomplete set of data~\cite{noei2019too}. To reduce such a bias on the findings, we collect all the reviews from December 2016 to April 2017 for the subject apps (1,722 reviews on average).

Third, the focus of our study is to examine the \textit{association} between performance costs of ads and the user concerns. Note that association does not imply causation. Furthermore, we do not have records of consecutive app versions, we do lot have a complete picture. This limitation is shared by previous studies~\cite{ruiz2014relationship,DBLP:journals/tse/BavotaVBPOP15,DBLP:conf/icsm/TianNLH15} that analyze relationships between app characteristics and user ratings. Still, these studies along with ours are the first steps towards understanding of the factors that impact user ratings/concerns. In future, more advanced statistical analysis, e.g., causality analysis~\cite{retherford1993statistical}, can also be employed.

Finally, to alleviate background noise and obtain reliable performance cost values, we measure 51 times for each app version and a total of more than 2,000 times for all the subject apps. The average results are utilized for our study.

\subsection{Usefulness of our findings}
We adopt the Technology Acceptance Model (TAM)~\cite{DBLP:journals/misq/Davis89}, the most influential models of technology acceptance~\cite{charness2016technology}, to analyze the usefulness of our findings. TAM summarizes two primary factors that can influence an individual's intention to adopt a technology: perceived ease of use (PEOU) and perceived usefulness (PU). Based on TAM~\cite{DBLP:journals/misq/Davis89}, the PEOU factor in our scenario could be affected by the developers' experience and voluntariness. We suppose that the developers are experienced in in-app ad design and voluntary to apply our findings to their practical development; so the PEOU factor is favorable for the usage of our findings. For the PE factor, it could be impacted by developers' subjective norm, their understanding of our findings, job relevance, expectation of higher quality, and demonstrability of the results, besides the PEOU factor. In the study, we have demonstrated that the obvious performance costs of in-app ads versions and the users' sensitivity to the performance costs through practical experiments, based on which we suppose that the developers believe our findings are meaningful and comprehend them well. We also assume that the developers do not refuse to try the findings to mitigate the performance costs of the in-app ads. The expected results can be better user experience or app revenue. Therefore, the PE factor would also be positive for the adoption of our findings, and the developers will have the attitude and intention to use the findings. However, the perception may change depending on age and gender~\cite{DBLP:journals/misq/Davis89}.


\subsection{\yun{Common ad-related terms in ad reviews}}
\yun{To take a deep look into what users commonly complain about ads, we use RankMiner to identify ``ad''-related terms and quantify user concern of each term. The ``ad''-related terms are determined by retrieving most similar terms to ``ad'' or ``ads'' following the method in Section~\ref{subsec:vis}. We find that users mentioned most about ad content (e.g., ``spam''), appearance style (``pop up ad''), ad size (e.g, ``full screen ad''), ad timing (e.g, ``30 second ads''), and obstruction (e.g., ``intrusive ad''). We manually label the ``ad''-related terms into these five groups, and visualize them for readers to better understand the extracted common ad-related complaints. We can discover that users are concerned about various aspects of advertising in apps besides the performance costs studied in this paper. Future research can extend our research by analyzing user perceptions of these aspects.}

\begin{figure}
	\centering
	\includegraphics[width=0.48 \textwidth]{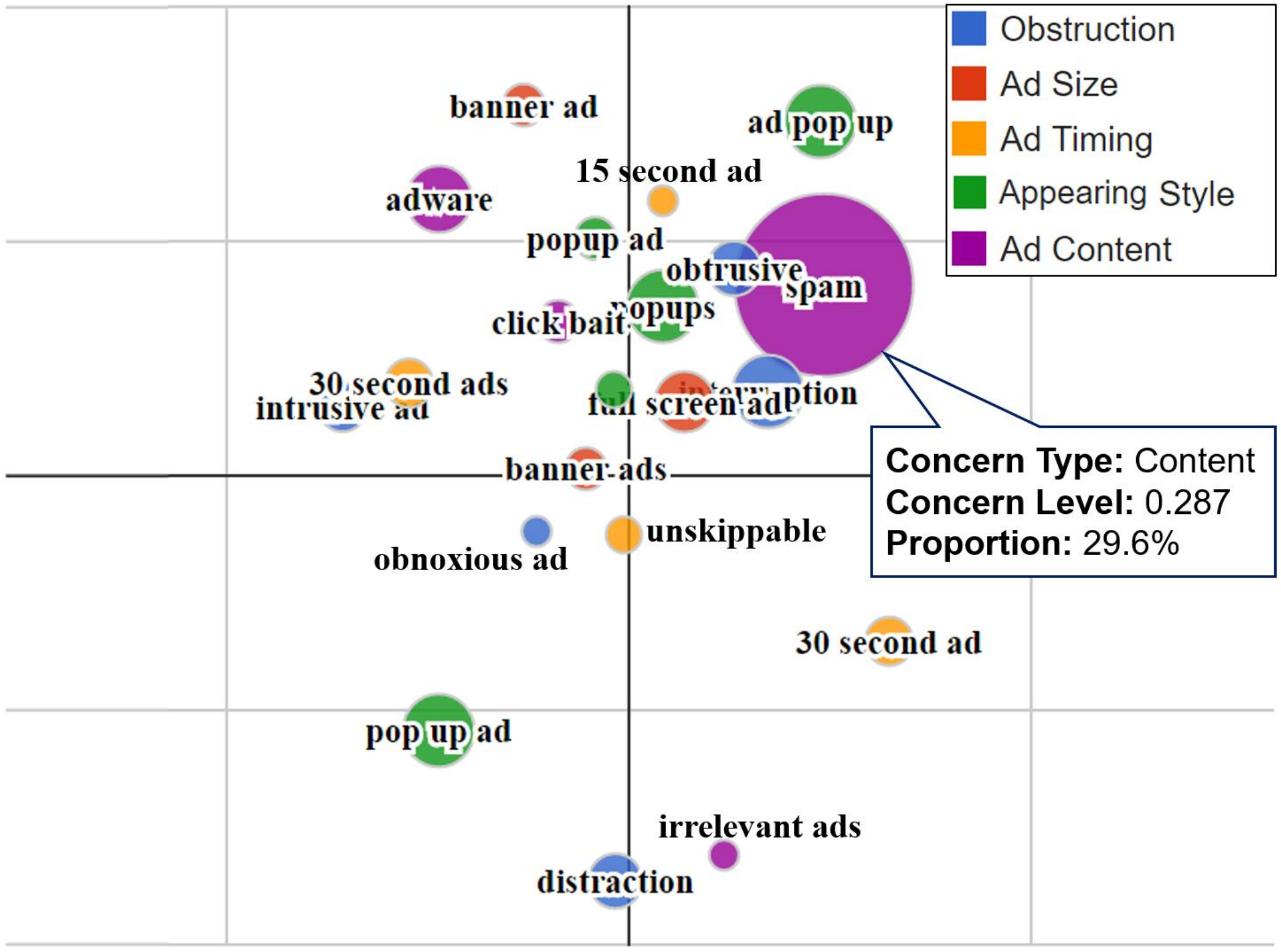}
	\caption{Visualization of ad issues. Larger bubbles indicate that the corresponding terms are of more concern to users.}
	\label{fig:adissue}
\end{figure}

\subsection{\add{Implications of our study}}
\add{\textbf{For practitioners:} The finding that performance costs of in-app ads versions are significantly larger than those of no-ads versions indicates that practitioners should notice the performance costs of in-app ads. The finding that users care most about the battery cost among all the performance cost types, suggests that practitioners should focus on the battery cost of in-app ads instead of treating all the cost types equally. To alleviate the negative impact of battery cost, practitioners can conduct A/B testing experiments to measure the battery cost of the in-app ads with different rendering strategies, e.g., different video resolutions and image sizes.}

\add{\textbf{For researchers:} More research on mitigating the costs of in-app ads including other hidden costs, such as app maintenance effort caused by in-app ads, is encouraged. Although anecdotal evidence exhibits the hidden costs of in-app ads, few research work has explored how to properly design mobile ads to mitigate the costs while preserving user experience (e.g., which rendering modes, such as image/video, of in-app ads are more favorable).}
\section{Related Work}\label{sec:literature}
We present two lines of work that inspire our study on in-app ads: app review analysis and ad cost exploration. A comprehensive survey on app store analysis for software engineering can be found elsewhere \citep{DBLP:journals/tse/MartinSJZH17}.

\subsection{App review analysis}

App review analysis explores the rich interplay between app customers and their developers. App reviews are a valuable resource provided directly by the users, which can be exploited by app developers during bug-fixing~\citep{DBLP:journals/tse/AliGA13,DBLP:conf/wcre/PelloniGCPPG18} and feature-improving process~\citep{paid2015cygao}.
In previous work~\citep{DBLP:conf/bcshci/IacobVH13}, the authors manually label 3,278 reviews of 161 apps, and discover the most recurring issues users report through reviews. Since mining app reviews manually is labor-intensive due to the large volume, more attempts on automatically extracting app features are conducted in prior studies. For example, \citet{DBLP:conf/msr/IacobH13} design MARA for retrieving app feature requests based on linguistic rules. 
Man \textit{et al.}~\citep{man2016experience} propose a word2vec-based approach for collecting descriptive words for specific features, where word2vec~\citep{DBLP:conf/nips/MikolovSCCD13} is utilized to compute semantic similarity between two words. Another line of work focuses on condensing feature information from reviews and captures user needs to assist developers in performing app maintenance~\citep{di2016would, villarroel2016release}. There are also investigations aiming at extracting valuable information from user reviews for supporting the evolution of mobile apps ~\citep{cuiyun2018idea,DBLP:journals/jss/PalombaVBOPPL18}. Previous research~\citep{vu2015mining,vu2016phrase}  has also investigated how to facilitate keyword retrieval and anomaly keyword identification by clustering semantically similar words or phrases.  

Other work \citep{guzman2014users, DBLP:conf/kbse/GuK15, DBLP:conf/www/LuizVAMSCGR18} propose methods to identify user opinions about specific app features/aspects. Detailed literature about opinion mining from app reviews can be found in the work by~\citep{DBLP:journals/jss/Genc-NayebiA17}. We use the sentiment prediction method proposed by Guzman \textit{et al.}~\citep{guzman2014users} for computing the sentiment score in RankMiner. Besides, the keyword extraction step in RankMiner builds on the work of Vu \textit{et al.}~\citep{vu2015mining} by extending the keyword lists with phrases instead of using single words only.




\subsection{Ad cost exploration}

Mobile ads can generate several types of costs for end users, \textit{e.g.}, battery drainage~\citep{DBLP:conf/wimob/MouawiECK15}, privacy leakage~\citep{DBLP:conf/wmcsa/ChenUKB14,DBLP:conf/ndss/RastogiSCPZR16,DBLP:conf/ndss/MengDCHL16}, and traffic data cost~\citep{DBLP:conf/imc/PujolHF15}. According to the research \citep{DBLP:journals/software/KhalidSNH15}, privacy \& ethics and hidden cost are the two most negatively perceived complaints (and are mostly in one-star reviews) among all studied complaint types. The work by Son \textit{et al.}~\citep{sooel2016what} shows that malicious ads can infer sensitive information about users by accessing external storage. Stevens \textit{et al.}~\citep{stevens2012investigating} investigate the effect on user privacy of popular Android ad providers by reviewing their use of permissions. The authors show that users can be tracked by a network sniffer across ad providers and by an ad provider across applications. The study by Gui \textit{et al.}~\citep{DBLP:conf/greens/GuiLWH16} proposes several lightweight statistical approaches for measuring and predicting ad related energy consumption, without requiring expensive infrastructure or developer effort. Wei \textit{et al.}~\citep{wei2012profiledroid} and Nath \textit{et al.}~\citep{nath2015madscope} discover that the ``free'' nature of apps comes with a noticeable cost by monitoring the traffic usage and system calls related to mobile ads. The work by Ullah \textit{et al.}~\citep{DBLP:conf/infocom/UllahBKK14} finds that although user's information is collected, the subsequent usage of such information for ads is still low.
Ruiz \textit{et al.}~\citep{ruiz2014impact} also explores how many ad libraries are commonly integrated into apps, and whether the number of ad libraries impacts app ratings. The authors find no evidence that the number of ad libraries in an app is related to its possible rating in the app store, but integrating certain ad libraries can negatively impact an app's rating.

To alleviate these threats, Mohan ~\citep{mohan2013prefetching} and Vallina-Rodriguez et. al~\citep{vallina2012breaking} develop a system to enable energy-efficient ad delivery. 
In the work of Seneviratne ~\citep{DBLP:conf/comsnets/SeneviratneTSKM13}, the authors propose the architecture MASTAds allowing ad networks to obtain only the necessary information in providing targeted advertisements with user privacy preserved. An interesting empirical study by Gui  \citep{gui2015truth} exhibits obvious hidden costs caused by ads from both developers' perspective (\textit{i.e.}, app release frequencies) and users' perspective (\textit{e.g.}, user ratings). Saborido ~\citep{DBLP:conf/iwpc/SaboridoKAG17} further highlight that ad-supported apps consume more resources than their corresponding paid versions with statistically significant differences. The work by Gao ~\cite{DBLP:conf/kbse/GaoZSLK18} investigates the performance costs raised by different advertisement schemes. In particular, they carried out an empirical study by considering 12 ad schemes from three different ads providers and analyzing three types of performance costs (memory/CPU, traffic and battery). The results of their study indicate that some ad schemes that produce less performance cost and provide suggestions to developers on ad scheme design.


In terms of performance cost measurement, the closest studies to our work are those by Gui ~\cite{gui2015truth} and Gao ~\cite{DBLP:conf/kbse/GaoZSLK18}. Different from them, we focus on analyzing the correlations between the performance costs of ads and users' attitudes. Besides, our performance costs are measured based on collected practical usage traces instead of experimental usage paths, which gives further confirmation on the findings by Gui ~\cite{gui2015truth}.

\section{Conclusion}\label{sec:conclusion}
In this paper, we have explored the effects of the performance costs of in-app ads on user experience. 
\add{We propose an approach, named RankMiner, for quantifying user concerns about app issues. The usefulness of RankMiner is embodied in that it can be beneficial for product managers to assess users' attitude towards specific app features and app testers to pinpoint possible app bugs based on the quantified user concerns. Besides, the deployment of RankMiner requires no professional knowledge about the involved techniques, reflecting its feasibility in practical technology transfer. In this work, we adopt RankMiner to measure user opinions about the performance costs of ads.}
We find that performance costs of with-ads versions are significantly larger than those of no-ads versions with negligible effect sizes.
By analyzing the correlations between the ads' performance costs and their impact on user opinions, we find the cost types that are more cared by users. We find that users are more concerned about the battery costs of ads, and tend to be insensitive to ads' data traffic costs. In future, we will extend our experiments by involving more apps, and study how to alleviate the battery costs when rendering ads.

\printcredits

\bibliographystyle{cas-model2-names}

\bibliography{cas-refs}









\end{document}